\documentclass[superscriptaddress,pre,twocolumn,amsmath,amssymb]{revtex4} 
\usepackage{graphicx,natbib}
\usepackage[caption=false]{subfig}

\newcommand{\vect}[1]{\mathbf{#1}} 
\newcommand{\RM }[1]{\mathrm{#1}}
\newcommand{\ave}[1]{ {\langle {#1} \rangle} }

\def\kB{ k_{\RM{B}} }

\def\Pc{{ \varphi }}
\def\DRosen{{D\rho^{1/3}(m/\kB T)^{1/2}}}

\def\br{\vect{r}}
\def\stwo{{s^{(2)}}}
\def\sx{{ s^{\RM{ex}} }}

\newcommand{\stwoi}[1][i]{{s^{(2)}_{#1}}}

\def\Dr{{D^{\text{R}}}}
\newcommand{\Dri}[1][i]{{D^{\text{R}}_{#1}}}
\def\Dz{{D^{\text{Z}}}}
\newcommand{\Dzi}[1][i]{{D^{\text{Z}}_{#1}}}
\def\Dgr{{D^{\text{GR}}}}
\newcommand{\Dgri}[1][i]{{D^{\text{GR}}_{#1}}}
\def\DgrC{{D^{\text{GR,C}}}}

\newcommand{\rhoDoi}[1][i]{{[\rho D_{#1}]_0 }}

\begin{document}

\title{Generalized Rosenfeld scalings for tracer diffusivities in 
not-so-simple fluids: \\
Mixtures and soft particles}

\author{William P. Krekelberg}
\email{wpkrekelberg@gmail.com}
\author{Mark J. Pond}
\author{Gaurav Goel}
\address{Department of Chemical
  Engineering, University of
  Texas at Austin, Austin, TX 78712.}

\author{Vincent K. Shen}
\email{vincent.shen@nist.gov}
\address{Chemical and Biochemical Reference Data Division, National
  Institute of Standards and Technology, Gaithersburg, Maryland,
  20899-8320, USA}

\author{Jeffrey R. Errington}
\email{jerring@buffalo.edu}
\address{Department of Chemical and Biological Engineering, 
  University at Buffalo, The State University of New York, 
  Buffalo, New York 14260-4200, USA}

\author{Thomas M. Truskett}
\email{truskett@che.utexas.edu}
\thanks{Corresponding Author} 
\address{Department of Chemical
  Engineering, University of
  Texas at Austin, Austin, TX 78712.}  
 \address{Institute for Theoretical Chemistry, 
   University of Texas at Austin, Austin, TX 78712.}

\begin{abstract}
  Rosenfeld [Phys. Rev. A \textbf{15}, 2545 (1977)] originally noticed that casting the transport
  coefficients of simple monatomic, equilibrium fluids in a specific
  dimensionless form makes them approximately single-valued functions
  of excess entropy.  This observation has predictive value because,
  while the transport coefficients of dense fluids can be difficult to
  estimate from first principles, the excess entropy can often be
  accurately predicted from liquid-state theory.  In this work, we use
  molecular simulations to investigate whether Rosenfeld's observation
  is a special case of a more general scaling law relating the tracer
  diffusivities of particles in mixtures to the excess entropy.
  Specifically, we study the tracer diffusivities, static structure,
  and thermodynamic properties of a variety of one- and two-component
  model fluid systems with either additive or non-additive
  interactions of the hard-sphere or Gaussian-core form.  The results
  of the simulations demonstrate that the effects of mixture
  concentration and composition, particle-size asymmetry and additivity,
  and strength of the interparticle interactions in these fluids are
  consistent with an empirical scaling law relating the excess entropy
  to a new dimensionless (generalized Rosenfeld) form of tracer
  diffusivity, which we introduce here.  The dimensionless form of the
  tracer diffusivity follows from knowledge of the intermolecular
  potential and the transport/thermodynamic behavior of fluids in the
  dilute limit.  The generalized Rosenfeld scaling requires less
  information, and provides more accurate predictions, than either
  Enskog theory or scalings based on the pair-correlation contribution
  to the excess entropy.  As we show, however, it also suffers from
  some limitations, especially for systems that exhibit significant
  decoupling of individual component tracer diffusivities.
\end{abstract}
\maketitle

\section{Introduction}
\label{sec:Introduction}

Many computational and experimental studies have now provided
empirical evidence of a strong correlation between transport
coefficients and the excess entropy of equilibrium fluids (see, e.g.,
\cite{Rosenfeld1977Relation-betwee,Rosenfeld1999A-quasi-univers,Dzugutov1996A-univeral-scal,Mittal2007Relationships-b,Abramson2007Viscosity-of-wa,Abramson2008Viscosity-of-ni,Abramson2009Viscosity-of-ca}).
The transport coefficients (e.g., diffusivity, viscosity, and thermal
conductivity) quantify the dynamic response of a fluid to a small
perturbation in the associated field variables, while the excess
entropy is a negative quantity that characterizes the number of
microstates rendered inaccessible to the fluid (relative to an ideal
gas) due to static interparticle correlations.  Changes to macrostate
variables that strengthen the interparticle correlations, and hence
make excess entropy more negative, typically result in slower
dynamical processes \cite{Rosenfeld1999A-quasi-univers}.  This is true
even for confined fluids
\cite{Mittal2006Thermodynamics-,Mittal2007Relationships-b,Mittal2007Does-confining-,Mittal2007Confinement-ent,Goel2009Available-state,Goel2008Tuning-Density-}
and for systems that show
anomalous dependencies of transport coefficients on density,
temperature, or the strength of the interparticle attractions 
\cite{Errington2001Relationship-be,Shell2002Molecular-struc,Truskett2002A-Simple-Statis,Kumar2005Static-and-dyna,Esposito2006Entropy-based-m,Netz2006Thermodynamic-a,Xu2006Thermodynamics-,Mittal2006Quantitative-Li,Errington2006Excess-entropy-,Sharma2006Entropy-diffusi,Oliveira2007Interplay-betwe,Yan2007Structure-of-th,Szortyka2007Diffusion-anoma,Oliveira2008Waterlike-hiera,Krekelberg2008Structural-anom,Yan2008Relation-of-wat,Krekelberg2007How-short-range,Krekelberg2009Anomalous-struc,Pond2009Composition-and,Chaimovich2009Anomalous-water}.

The connection between transport coefficients and excess entropy is of
fundamental interest because it provides a clue in the long-standing
puzzle concerning what structural and thermodynamic properties
correlate with the dynamics of equilibrium fluids.  The link also has
practical consequences.  For example, if the transport coefficients of
a fluid, cast in an appropriately reduced form, can be approximately
represented as a single-valued function of the excess entropy, then
knowledge of the latter allows indirect ``prediction'' of the former \cite{Goel2009Available-state}.
The value of this approach lies in the fact that, while it is
difficult to directly estimate transport coefficients from first
principles, the excess entropy can often be accurately predicted from
liquid-state theories.

At present, a rigorous and general statistical mechanical
justification for the empirically observed relationship between
transport coefficients and excess entropy is lacking.  However, even
in the absence of a formal justification, there are a number of
practical questions that deserve further investigation.  Here, we
present new calculations that address key aspects of two such
questions.
\begin{itemize}
\item For what types of fluid systems is tracer diffusivity, when cast in
    an appropriately reduced form, approximately a single-valued
    function of excess entropy?
\item Can we develop a strategy for determining the
  aforementioned ``appropriately reduced form'' for tracer diffusivity
  of a given system from knowledge of the intermolecular potential,
  temperature, and composition?
\end{itemize}

To understand the context of these questions, it is helpful to first
consider some background.  It has long been appreciated that the
following reduced form of self-diffusivity, $\Dr\equiv\DRosen$, can be
formally represented as a single-valued function of excess entropy,
$\sx$, for any model fluid of identical particles which interact via
an inverse-power-law (IPL) pair potential of the form
$v(r)=\epsilon(\sigma/r)^n$ (see, e.g., \cite{Hoover1991Computational-S}).  Here, 
$D$ is self-diffusivity, $\rho$ is number density, $T$ is temperature,
$k_\text{B}$ is the Boltzmann constant, $m$ is the particle mass, and
the combination $\epsilon \sigma^n$ is the single parameter of the IPL
potential.  The function relating $\Dr$ and $\sx$ strictly depends on
the exponent $n$, but the dependence is weak.  In fact, Rosenfeld \cite{Rosenfeld1999A-quasi-univers}
first pointed out that the relationship is ``quasi-universal'' in the sense that,
for a given value of excess entropy, there is less than 30\% variation
in the predicted self-diffusivities for different equilibrium IPL
fluids with $4\le n \le\infty$.  Because of this observation, we refer
to $\Dr$ in this work as the Rosenfeld form of reduced diffusivity.
Recently, Mittal et al. \cite{Mittal2007Relationships-b} demonstrated that the same quasi-universal
relationship also adequately describes the correlation between $\Dr$
and $\sx$ for both bulk (isotropic) and confined (inhomogeneous)
equilibrium Lennard-Jones fluids.  An important point of the Rosenfeld \cite{Rosenfeld1999A-quasi-univers}
and Mittal et al. \cite{Mittal2007Relationships-b} investigations, discussed from a different angle in
a more recent study by Dyre and co-workers \cite{Gnan2009Pressure-energy}, is that since both the
static and dynamic properties of many dense, simple liquids are
dominated by their repulsive interactions, they closely mimic the
behaviors of ``equivalent'' IPL fluids.  Thus, one can expect $\Dr$ of
these monatomic simple-liquid systems to scale with $\sx$ in a way
that is consistent with the trend originally identified by Rosenfeld.

Yet there are many fluids that cannot be expected to mimic the
behavior of monatomic IPL systems.  Will excess entropy also prove
useful for predicting the dynamics of these more complex fluids?  For
example, can excess entropy be used to reliably predict the effects of
temperature and density on the self-diffusivity of model fluids with
soft (or even bounded) pair potentials, like those that characterize
the effective interactions between macromolecules or micelles in
solution \cite{Lang2000Fluid-and-solid,Louis2000Mean-field-flui,Likos2001Effective-inter}?
  At present the answer is unclear.  A recent molecular
dynamics simulation study by \citet{Krekelberg2009Anomalous-struc} demonstrated that $\Dr$
of the Gaussian-core fluid is not (even approximately) a single-valued
function of $\sx$, and the same is true for the Rosenfeld-scaled thermal
conductivity and viscosity~\cite{Mausbach2009Transport-}.  Is there a systematic way to construct, based on
knowledge of the pair potentials and temperature of a system, an
alternative reduced form, i.e. a generalized Rosenfeld diffusivity
$\Dgr$, that (to within acceptable tolerances) is a function of excess
entropy alone?  One of the goals of this study is to address this
question for different types of model systems with a variety of
interactions, [e.g., continuous or discontinuous, steeply repulsive
(diverging) or soft (bounded)].

A related question is whether excess entropy can be used to predict
the effects of temperature, density, and composition on the tracer
diffusivities of the components of a fluid mixture?  This question has
been recently studied in a limited context.  Specifically, following
initial work on monatomic systems by
\citet{Dzugutov1996A-univeral-scal}, a scaling for the tracer
diffusivities of mixtures based on two-body contributions to the
excess entropy has been introduced \cite{Samanta2001Universal-Scali}.
Although it appears that this mixture version of the Dzugutov scaling
can capture some of the behaviors exhibited by simple fluid systems,
it also has some significant limitations.  For example, the reduced
diffusivity for the Dzugutov scaling, $\Dz$, relies on defining an
effective ``hard-core diameter'' for each interparticle potential,
which is not convenient for the study of soft or penetrable particles
with bounded interactions.  Moreover, the Dzugutov scaling fails to
describe the behavior of systems in the limit of vanishing number density.
Finally, computing the two-body excess entropy requires knowledge of
the radial distribution functions between all components in the
mixture for each thermodynamic state of interest, which is
particularly cumbersome when studying inhomogeneous fluids.  This
should be contrasted with the excess entropy used in the Rosenfeld
scaling, which can be readily calculated from knowledge of the fluid's
equation of state.  For all of these reasons, we examine in this study
whether one can, based on knowledge of mixture composition,
temperature, and pair potentials, construct a generalized Rosenfeld
form for the reduced tracer diffusivity for component $i$, $\Dgri$,
that is approximately a single-valued function of the excess entropy
of the fluid mixture.

The organization of the paper is as follows.  In
Section~\ref{sec:thermo_transport}, we introduce the simple idea that
underlies the generalized Rosenfeld scaling for predicting tracer
diffusivity from excess entropy.  Section~\ref{sec:methods} provides
details on the model fluid systems and the simulation techniques that
we use here to put the predictions of this scaling to quantitative
tests.  In Section~\ref{sec:results_discussion}, we analyze the
generalized Rosenfeld scaling for a wide variety of binary mixtures of
hard spheres, Widom-Rowlinson particles, and Gaussian-core particles.
In Section~\ref{sec:conclusions}, we discuss how this data 
helps to understand the strengths and
limitations of using excess entropy for predicting the effects that
macroscopic parameters (temperature, density, composition) and
microscopic details (particle diameter, particle mass, softness of the
interparticle potential) have on single-particle dynamics.

\section{Generalized Rosenfeld form for Reduced Tracer Diffusivity}
\label{sec:thermo_transport}

In order to ensure that $\Dgri$, the generalized Rosenfeld form of the
reduced tracer diffusivity of component $i$, is defined in a way that
is useful for making excess-entropy based predictions, we aim to have
it satisfy the following three criteria for a given system.  (i) It
should be proportional to the bare tracer diffusivity, i.e., $\Dgri=
\alpha D_i$.  (ii) The prefactor $\alpha$, which has units of
reciprocal diffusivity, should be readily calculable based on
knowledge of the parameters that define the fluid system, i.e.,
macroscopic variables like temperature $T$, density $\rho$, and the
mole fractions of the species, as well as microscopic parameters like
particle masses and the pair potentials $V_{ij}(r)$ describing the
effective interactions.  (iii) The dimensionless quantity $\Dgri$ should
be approximately a single-valued function of $\sx$.

For the case of a monatomic IPL fluid, the aforementioned criteria are
rigorously satisfied at all state points if one adopts the Rosenfeld
form of the reduced self-diffusivity $\Dr$ discussed in the
introduction.  In fact, Rosenfeld \cite{Rosenfeld1999A-quasi-univers} previously illustrated that, at low
number density, an analytical relationship between $\Dr$ and $\sx$ can
be obtained for an IPL fluid by using two equations: an Enskog theory
expression for $D$ and a truncated second-virial expansion for $\sx$.

Our approach here is to similarly examine the low-density limit of
more complex model fluids and mixtures, taking advantage of known
theoretical results for the tracer diffusivity and excess entropy to
seek out a potentially useful definition for $\Dgri$.  As we show
below, these low-density theoretical results do suggest a simple
expression for $\Dgri$ that, in the dilute limit, satisfies the three
criteria mentioned above.  Of course, unlike for an IPL fluid,
choosing $\Dgri$ of a complex fluid so that it is a single-valued
function of $\sx$ at low density does not guarantee that it will also
behave that way at high particle density.  In fact, one can view the
extent to which the $\Dgri$ versus $\sx$ scaling holds at higher
particle density as a measure of the utility of excess entropy for
predicting tracer diffusivity of a given fluid system.
Section~\ref{sec:results_discussion} focuses on quantitatively
examining this point for a variety of model fluids.

To make the above discussion more concrete, first consider that kinetic
theory indicates that $D_i$ is inversely proportional to number
density $\rho$ in a
fluid mixture at low density~\cite{footnote1}
\begin{equation}
  \label{eq:D_low_density}
  D_i=\frac{\rhoDoi}{\rho},
\end{equation}
where the quantity $\rhoDoi\equiv \lim_{\rho\rightarrow 0} \rho
D_i$ generally depends on temperature, mixture composition, as well as
the masses of the species and their
the interparticle interactions.  We discuss simple theoretical 
methods for
estimating $\rhoDoi$ for model systems below.  The 
relationship between $\rho$ and $\sx$, to leading order in $\rho$, can
be expressed as
\begin{equation}
  \label{eq:sx_second_virial}
  \sx/\kB=-\rho\left[ B+T\frac{d B}{d T} \right],
\end{equation}
where $B$ is the second virial coefficient, given by
\begin{subequations}
  \label{eq:B2}
  \begin{equation}
    \label{eq:B2mix}
    B=\sum_{i}^{N_c} \sum_{j}^{N_c} x_i x_j B_{ij}.
  \end{equation}
  Here, the sums are over the $N_c$ components of the mixture, $x_i$ is the
  mole fraction of component $i$, and $B_{ij}$ can be expressed in
  terms of the pair potentials $V_{ij}(r)$ as
  \begin{equation}
    \label{eq:Bij}
    B_{ij}=2 \pi \int_0^{\infty} \left[ 1-e^{-\beta V_{ij}(r)} \right]
    r^2 dr,
  \end{equation}
\end{subequations}
where $\beta^{-1}=\kB T$.
Using Eq.~\eqref{eq:sx_second_virial} to eliminate $\rho$ from
Eq.~\eqref{eq:D_low_density} and rearranging leads to
\begin{equation}
  \label{eq:D_basic_scaling}
  \frac{D_i}{\left(B+T\frac{d B}{d T}\right)\rhoDoi}= \frac{1}{-\sx/\kB},
\end{equation}
which again is valid only in the the $\rho \rightarrow 0$ limit.  
We identify the dimensionless quantity on the left-hand-side, which is
clearly a function of $\sx$ only at low density, as the generalized Rosenfeld
reduced form of the tracer diffusivity, $\Dgri$, 
\begin{equation}
  \label{eq:def_D_generalized_rosenfeld}
  \Dgri\equiv\frac{D_i}{\left(B+T\frac{d B}{d T}\right)\rhoDoi}.
\end{equation}
Note that the expression for $\Dgr$, the
generalized Rosenfeld self-diffusion coefficient 
for a monatomic fluid, is obtained by
replacing $D_i$ with $D$ in Eq.~\eqref{eq:def_D_generalized_rosenfeld}.  

Although the definition for $\Dgri$ given in
Eq.~\eqref{eq:def_D_generalized_rosenfeld} is compact, it is more
convenient for making predictions for model systems if $\rhoDoi$ is further 
expressed in terms of the mole fractions of the species, the associated 
pair potentials, the particle masses, and the temperature.  
Below, we present simple theoretical expressions for carrying this out
for particles with hard-sphere and soft (continuous) interactions, 
respectively.

\subsection{Hard-Particle Interactions}
\label{sec:hard-systems}

For models with hard-sphere interactions, an expression for $\rhoDoi$ is easily
obtained within Enskog kinetic theory \cite{Chapman1970The-Mathematica,Jacucci1975Structure-and-d}
In particular, the product $\rho D_i$ is
given by
\begin{equation}
  \label{eq:rhoD_enskog_HS}
  \rho D_i=\frac{3}{8 \pi^{1/2}}\frac{\sqrt{\kB T/m_i}}{ \sum_{j=1}^{N_c} x_j
    \sigma_{ij}^2 g(\sigma_{ij}^{+}) \left[\frac{1}{2}\left(1+\frac{m_i}{m_j}\right)\right]^{-1/2} },
\end{equation}
where $m_i$ is the mass of component $i$, $\sigma_{ij}$ is the
hard-sphere contact diameter between particles of type $i$ and $j$, and
$g_{ij}(\sigma_{ij}^{+})$ is the value of the radial distribution
function between particles of type $i$ and $j$ at contact. The low
density limit $\rhoDoi$ is obtained by substituting
$g_{ij}(\sigma_{ij}^{+})=1$ 
into Eq.~\eqref{eq:rhoD_enskog_HS}, which gives
 \begin{equation}
  \label{eq:rhoD0_enskog_HS}
  \rhoDoi=\frac{3}{8 \pi^{1/2}}\frac{\sqrt{\kB T/m_i}}{ \sum_{j=1}^{N_c} x_j
    \sigma_{ij}^2 \left[\frac{1}{2}\left(1+\frac{m_i}{m_j}\right)\right]^{-1/2} }.
\end{equation} 
When computing $\Dgri$ for the hard-sphere and
Widom-Rowlinson model mixtures discussed in Section~\ref{sec:methods}, we 
simply substitute Eq.~\eqref{eq:rhoD0_enskog_HS}
into Eq.~\eqref{eq:def_D_generalized_rosenfeld}.

\subsection{Soft-Particle Interactions}
\label{sec:soft-systems}
We also study fluids of soft particles in this work, i.e., particles
with continuous and bounded interactions that cannot be treated as
hard spheres with an effective temperature-dependent diameter.  In
order to predict $\rhoDoi$ for these models, we use an approximate
theory due to Tankeshwar and
co-workers~\cite{Tankeshwar1992Tracer-diffusio,Sharma1996Self-diffusion-},
which we refer to as the Tankeshwar diffusion model (TDM).  We have
found that this basic theoretical approach strikes a reasonable
balance between simplicity and accuracy.  It has been shown to
semi-quantitatively describe how temperature, composition, and density
affect the tracer diffusivity of Lennard-Jones fluids, the
one-component plasma, and Yukawa fluids \cite{Tankeshwar1991A-simple-model-}.  We have also found that it
approximately captures how temperature and density affect the
diffusion coefficient of the Gaussian-core fluid.

The details of the TDM are discussed extensively elsewhere \cite{Tankeshwar1991A-simple-model-,Tankeshwar1992Tracer-diffusio,Sharma1996Self-diffusion-}.  In short,
it is based on an approximate expression for the velocity
autocorrelation function, and hence the tracer diffusivity $D_i$ via
the Green-Kubo relation, for each component $i$ of the fluid in terms
of two parameters: the Einstein frequency $\omega_i$ and a ``jumping''
frequency $\tau_i$.  The values of these parameters are obtained by
ensuring that the velocity autocorrelation functions satisfy some
exact microscopic sum rules.

Within this model, the product $\rho D_i$ is given by
\begin{equation}
  \label{eq:rhoD_Einstein}
  \rho D_i= \rho \frac{\kB
    T}{m_i} \frac{\pi}{2} \tau_i \sec \left( \frac{\pi}{2} \omega_i \tau_i\right),
\end{equation}
where
\begin{equation}
  \label{eq:tau_omega_A2_A4}
  \begin{split}
    \tau_i^{-2}&=\frac{\rho A_i^{(4)}-\left[\rho A_i^{(2)}\right]^2}{4 \rho A_i^{(2)}}, \\
    \omega_i^{2}&=-\frac{5\left[\rho A_i^{(2)}\right]^2-\rho A_i^{(4)}}{4 \rho
      A_i^{(2)}},
  \end{split}
\end{equation}
and
\begin{subequations}
  \label{eq:A2_A4}
  \begin{align}
    A_i^{(2)}&=\frac{4\pi}{3}\sum_{j=1}^{N_c} \frac{x_j}{m_j}
    \int_0^{\infty} 
    dr\, r^2 g_{ij}(r)
    \left(\frac{2}{r} \frac{d V_{ij}}{dr}+\frac{d^2
        V_{ij}}{dr^2}\right), \label{eq:A2}\\
    A_i^{(4)}&=\frac{4\pi}{3} \sum_{j=1}^{N_c} \frac{x_j}{m_i}
    \left(\frac{1}{m_i}+\frac{1}{m_j}\right)
    \int_0^{\infty} dr\, r^2 g_{ij}(r)\nonumber\\ 
    &\qquad \qquad\times
    \left\{ \frac{2}{r^2}\left(\frac{d V_{ij}}{dr}\right)^2+
      \left(\frac{d^2 V_{ij}}{dr^2}\right)^2\right\}. \label{eq:A4}
  \end{align}
\end{subequations}
In Eq.~\eqref{eq:A2_A4}, $V_{ij}$ is the pair potential between
particles of species $i$ and $j$, and terms involving three-body static
correlations have been omitted.  In order to evaluate $\rhoDoi$, we
take the $\rho \rightarrow 0$ limit of Eqs.~\eqref{eq:rhoD_Einstein}--\eqref{eq:A2_A4},
which leads to 
\begin{equation}
  \label{eq:rhoD0_einstein}
  \rhoDoi=\frac{\kB T}{m_i} \left(\frac{A_{i,0}^{(4)}}{[A_{i,0}^{(2)}]^3}\right)^{1/2}.
\end{equation}
Here $A_{i,0}^{(2)} \equiv \lim_{\rho\rightarrow 0} A_{i}^{(2)}$ and
$A_{i,0}^{(4)} \equiv \lim_{\rho\rightarrow 0} A_{i}^{(4)}$, each of
which follow by replacing $g_{ij}(r)$ in Eq.~\eqref{eq:A2_A4}, with the
Boltzmann factor of the pair potential,
\begin{equation}
  \label{eq:pcf_low_dens}
 \lim_{\rho\rightarrow 0} g_{ij}(r)=\exp [-\beta V_{ij}(r)]. 
\end{equation}
When computing $\Dgri$ for the Gaussian-core mixtures discussed in 
Section~\ref{sec:methods}, we 
substitute Eq.~\eqref{eq:rhoD0_einstein}
into Eq.~\eqref{eq:def_D_generalized_rosenfeld}.

\section{Model Systems and Simulation Methods}
\label{sec:methods}

As discussed in Section~\ref{sec:Introduction}, a key aim of this
study is to investigate whether it is possible to construct an
excess-entropy based strategy for predicting tracer diffusivity
generic enough to be successfully applied to fluid mixtures with
either hard (impenetrable) or soft (penetrable) interparticle
interactions.  For our model systems, we choose familiar
representations for both: the hard-sphere (HS) pair potential for the
former and the Gaussian-core pair potential for the latter.

The HS pair potential is discontinuous and athermal, assigning
infinite energy to configurations that have particle overlaps
and zero energy to all others.  It is thus represented as
\begin{equation}
  \label{eq:HS}
  V^{\RM{HS}}_{ij}(r)=
  \begin{cases}
    \infty& r<\sigma_{ij}, \\
    0 & r\geq \sigma_{ij}.
  \end{cases}
\end{equation}
Here, $\sigma_{ij}$ is the contact diameter between particles of type
$i$ and $j$.  We investigate several binary HS mixtures with additive
diameters, i.e., $\sigma_{ij}=(\sigma_i+\sigma_j)/2$.  In particular,
we first study compositional effects on tracer diffusivity and excess
entropy using a system composed of equimass ($m_1/m_2=1$) particles
with diameter ratio $\sigma_1/\sigma_2=1.3$.  For this system, we
examine mole fractions of component one in the range $0.1 \le x_1 \le
0.9$.  We also investigate the effects of diameter ratio 
by considering particles with
$\sigma_1/\sigma_2=1.3,2.0,3.0,5.0$.  These
latter studies are carried out at fixed composition $x_1=0.1$. All of the
above systems are studied across a wide range of packing fractions
$\Pc=\pi(x_1\sigma_1^3+x_2\sigma_2^3)/6$ spanning between $0.05$
(dilute gas) and $0.5$ (near the single-component HS freezing transition).

We also consider a highly non-additive version of the binary HS
mixture: the Widom-Rowlinson (WR) model
\cite{Widom1970New-Model-for-t}.  In this system, the contact diameter
between particles of the same type is zero
($\sigma_{11}=\sigma_{22}=0$), but the cross-diameter is finite
$\sigma_{12}=\sigma_{21}=\sigma$ and $m_1=m_2$.  As might be imagined,
this system exhibits entropically driven phase separation at
sufficiently high density, which we avoid here by studying $0 <
\rho\sigma^3 < 0.7$.  Since the model is symmetric with respect to the
interactions, we can deduce global behavior by studying mole fractions
in the range $0 < x_1 \le 0.5$.

Finally, we study 
fluids composed of soft particles that interact via the 
bounded Gaussian-core pair potential \cite{Stillinger1978Study-of-meltin}, given by 
\begin{equation}
  \label{eq:GC_potential}
  V^{\RM{GC}}_{ij}(r)=\epsilon_{ij} \exp[-(r/\sigma_{ij})^2],
\end{equation}
where $\epsilon_{ij}$ and $\sigma_{ij}$ are parameters that
characterize the energy and length scale, respectively, of the
interaction between particles of type $i$ and $j$.  For the
simulations in this work, we 
truncate the interparticle 
interactions at a separation of 3.2$\sigma_{ij}$.  We examine both
single-component and two-component Gaussian-core fluids. For the latter, we
adopt the same parameters used in a previous investigation of the
static structure and thermodynamics of that
system~\cite{Archer2001Binary-Gaussian}.  Specifically,
we assign $\sigma_{22}=0.665 \sigma_{11}$ and
$\sigma_{12}=\sqrt{(\sigma_{11}^2+\sigma_{22}^2)/2}$,
$\epsilon_{11}=\epsilon_{22}$ and $\epsilon_{12}=0.944 \epsilon_{11}$
(which encourages mixing), and $m_1=m_2$.
We investigate the binary Gaussian-core fluid at compositions $0.1 \le x_1 \le 0.9$,
temperatures  $0.05 \le \kB T/\epsilon_{11} \le 0.4$, and
densities $0.05 \le \rho \sigma_{11}^3 \le 1.0$.  At some
temperatures, the maximum density in this range is not an isotropic 
fluid due to propensity of the system to phase separate.  
We excluded from our analysis any state points that
showed thermodynamic or structural indications of phase separation.

To explore the dynamic properties of the above systems, we perform
molecular dynamics (MD) simulations.  For the HS and WR mixtures, we use a
standard event-driven algorithm \cite{Rapaport2004The-Art-of-Mole}.
For the binary Gaussian-core fluid, the equations of motion are 
integrated using the velocity-Verlet method
\cite{Allen1987Computer-Simula} with time step
$\delta t=0.05 \sqrt{m_1 \sigma_{11}^2 / \epsilon_{11}}$.  All MD
simulations are carried out in the microcanonical ensemble with
$N=3000$--$5000$ particle using 
a periodically-replicated cubic simulation cell with volume
$V$, chosen in accord with the desired fluid density.  
Tracer diffusion coefficients $D_i$ are calculated by
fitting the long time average mean squared displacement of the $i$
type particles $\ave{ \delta r^2_i}$ to the Einstein relation $6D_i t=
\ave{\delta r^2_i}$.  Note that, for the case of the monatomic fluid,
this definition of the tracer diffusivity reduces to the
self diffusivity.  We perform multiple independent simulations at
several state points for each model, 
and we find the relative
standard error in tracer diffusivities to be less than 1\%.

Thermodynamic properties of the Gaussian-core fluid mixtures 
are computed using grand-canonical
transition-matrix Monte Carlo (GC-TMMC) simulations.  These simulations are
conceptually equivalent to a series of semigrand simulations performed
over a range of fluid densities stitched together using ghost
insertion/deletion moves.  Details of this method can be found
elsewhere \cite{Shen2006Determination-o,Shen2005Determination-o}. These
simulations require fixed values of the activity, $\{\xi_1,\xi_2\}$,
volume, $V$, and temperature, $T$, as inputs. The activity is defined
as $\xi_i=\Lambda_i^{-3}\exp(\mu_i/\kB T)$ , where $\mu_i$ and
$\Lambda_i$ are the chemical potential and the thermal de Broglie
wavelength of component $i$, respectively.
All GC-TMMC simulations for the Gaussian-core fluid mixtures reported here use 
a system volume of $V = 343$. 
For the activities, the values of $\ln{ \xi_i }$ that we use span from $32.63$ 
at the lowest temperature to $12.24$ at the highest temperature
investigated.
Thermodynamic properties at other values of activity are
obtained via the histogram re-weighting technique. The primary
quantity obtained from GC-TMMC is the particle number probability
distribution $\Pi(N;\{\mu_i\},V,T)$.  From this, excess entropies are
trivially calculated (see \cite{Mittal2007Confinement-ent}).  System
size effects in excess entropy for the Gaussian-core mixtures are found
to be negligle by comparing results to a series of simulations 
using a smaller volume of $V = 216$. We confirmed that 
the equation of state (and hence excess entropy) of Gaussian-core fluid
mixtures produced from
the GC-TMMC simulations is statistically 
indistinguishable with that produced from
molecular dynamics simulations.

GC-TMMC calculations for the WR mixtures are    
performed using a system volume of $V = 343$. For this fluid,
simulations are completed with activity values of $\xi_1 = \xi_2 = 1$,
and histogram re-weighting is applied to obtain thermodynamic
quantities at other values of activity. 
System size effects are examined by performing simulations over a
limited density and composition range with a volume of $V = 1000$, and
are also found to be negligible.

The excess entropy data we present for the binary HS mixtures
is calculated from the accurate 
Boublik-Mansoori-Carnahan-Starling-Leland (BMCSL) equation of 
state~\cite{Boublik1970,Monsoori1971}.
As a check, we compared the BMCSL values for compressibility
factor and excess entropy against those obtained via molecular dynamics
simulations for selected state points as a function of particle
diameter ratio and packing fraction, and we found the agreement  
to be excellent.

Finally, we also compare the results for the binary systems to corresponding
single-component systems.  For the single-component HS system we use
the data of \cite{Goel2009Available-state} and for the
single-component Gaussian-core fluid we use the data of \cite{Krekelberg2009Anomalous-struc}.

\section{Results and discussion}
\label{sec:results_discussion}



\subsection{Hard-sphere mixtures}
\label{sec:hs-mixtures }

\subsubsection{Compositional effects}
\label{sec:hs_comp_effects}

\begin{figure*}
  \mbox {
    \includegraphics[clip]{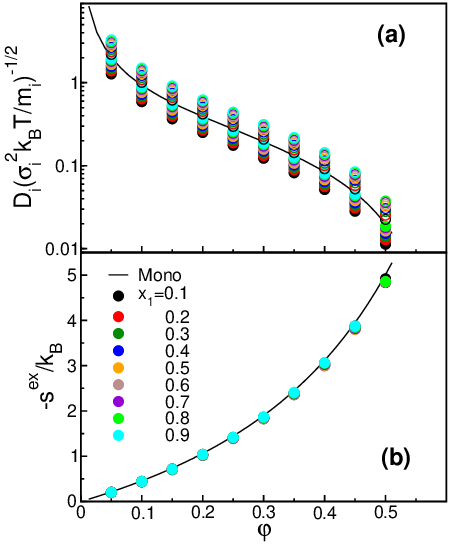}
    \includegraphics[clip]{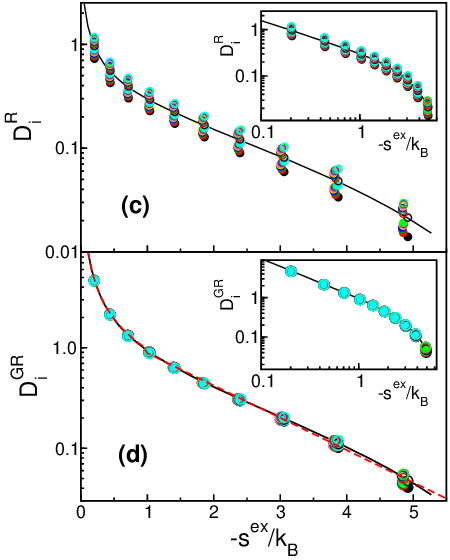}
  }

  \subfloat{\label{fig:HSB_D_vs_phic_comp}}
  \subfloat{\label{fig:HSB_sx_vs_phic_comp}}
  \subfloat{\label{fig:HSB_DRi_sx_comp}}
  \subfloat{\label{fig:HSB_DSi_vs_sx_comp}}
  \caption{ (color online) Properties of the binary HS mixture with
  particle diameter ratio $(\sigma_1/\sigma_2)=1.3$, equal mass, and a variety of 
compositions. (a)
    Tracer diffusion coefficients $D_i$ and (b) (negative) excess entropy $-\sx$ as
    a function of packing fraction $\Pc$.  (c) Rosenfeld
    $\Dri$ and (d) generalized Rosenfeld $\Dgri$ tracer diffusivities 
as a function of
    $-\sx$.  Filled and open symbols denote component $1$ (large) and
    $2$ (small),
    respectively.  The color of symbols denotes the mole fraction of
    component $1$, $x_1$, specified in the legend of (a).  The solid line in
    each figure is the result for the single-component HS system.  The
    dashed red line in (d) represents a least-squares fit of the data
    to Eq.~\eqref{eq:Dgri_fit}, which results in $\alpha=0.95$,
    $A=1.85$, and $\beta=0.74$. In (c) and (d), the insets are the
    same as the main plots, but on a log-log scale.  }
  \label{fig:HSB_scalings_comp}
\end{figure*}

We begin by investigating the effects of composition on mixtures of
HS particles with size ratio $\sigma_1/\sigma_2=1.3$.  
Figure~\ref{fig:HSB_D_vs_phic_comp} displays the
tracer diffusion coefficients $D_i$ of the two 
components as a function of total
packing fraction $\Pc$ for several different compositions, 
indicated by the mole fraction of large particles, $x_1$.  
As must be the case, when one of the species is present in high
concentration, its tracer diffusivity approaches the value of 
the self-diffusion coefficient~$D$ of the single-component HS fluid 
at the same
packing fraction $\Pc$.  
However, the tracer diffusivity of the dilute component is generally 
different than $D$.  In particular, when component $1$ (the larger
particles) is dilute, one should expect $D_1<D$.  This
logic can be qualitatively rationalized by the fact that, on average, 
motion of the
larger solute would require larger local structural rearrangements (i.e.,
fluctuations) than for the 
motion of the smaller solvent particles.  Conversely,
by an analogous argument, one expects $D_2>D$ if component $2$ 
(the smaller particles) is dilute. The data in
Figure~\ref{fig:HSB_D_vs_phic_comp} is consistent with these
expectations. 

It is interesting to note that the compositional variation in 
$\ln D_i$ is fairly insensitive to the value of 
$\Pc$.  Moreover, the
excess entropy [Figure~\ref{fig:HSB_sx_vs_phic_comp}] exhibits almost
no compositional dependence whatsoever, and its packing fraction
dependence for any particular composition 
is nearly identical to that of the single-component HS fluid.  All of
this suggests that an appropriate 
composition dependent rescaling of the
tracer diffusivity data might (approximately) make it a single-valued
function of excess entropy.  Indeed, 
Figures~\ref{fig:HSB_DRi_sx_comp} and \ref{fig:HSB_DSi_vs_sx_comp}
show that the generalized Rosenfeld tracer diffusivities $\Dgri$ of 
Eq.~\ref{eq:def_D_generalized_rosenfeld} for both species 
collapse onto a single curve [that describing 
the single-component (SC) fluid data,
$D^{\RM{GR}}_{\RM{SC}}$] when
plotted versus excess entropy, while no data collapse occurs if the 
tracer diffusivities are na\"{i}vely represented in the original 
Rosenfeld reduced form $\Dri$.

The single-component relationship
$D^{\RM{GR}}_{\RM{SC}}(\sx)$ can be described by a piece-wise
function.  The form of its low-density (low $-\sx$) scaling is an inverse power
law, given by
Equation~\eqref{eq:D_basic_scaling}. From
Figure~\ref{fig:HSB_DSi_vs_sx_comp}, we infer that when
$-\sx/\kB\gtrsim 1$,
the relationship becomes approximately exponential.  A least squares fit
assuming these generic function forms, i.e.,
\begin{equation}
  \label{eq:Dgri_fit}
  D^{\RM{GR}}_{\RM{SC}}(\sx) =
  \begin{cases}
    \alpha[-\sx/\kB]^{-1}& -\sx/\kB<1,\\
    A\exp[-B \sx/\kB] & -\sx/\kB>1,
  \end{cases}
\end{equation}
yields $\alpha=0.95$, $A=1.85$, and $\beta=0.74$, and describes 
the simulation data very well (see
red dashed line in Fig.~\ref{fig:HSB_DSi_vs_sx_comp}).

\begin{figure}[h]
  \centering
  \includegraphics[clip]{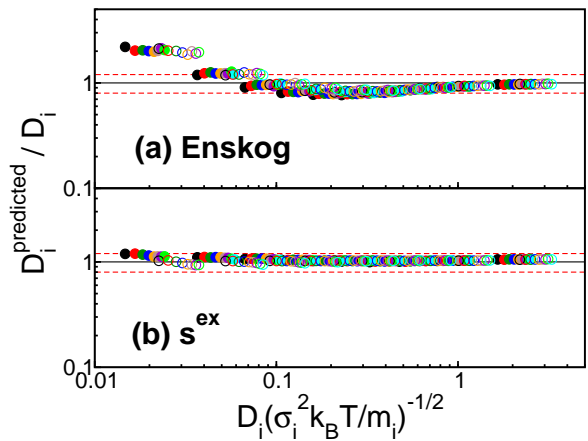}
  \subfloat{\label{fig:HSB_comparison_comp_enskog}}
  \subfloat{\label{fig:HSB_comparison_comp_sx}}
  \caption{(color online) Ratio of tracer diffusivity predicted from (a)
    Enskog theory [Eq.~\eqref{eq:rhoD_enskog_HS}] and (b) from excess
    entropy and the single-component HS result (generalized Rosenfeld scaling) [Eq.~\eqref{eq:Dpredicted_sx}], for a HS mixture with
    $(\sigma_1/\sigma_2)=1.3$, equal mass, and a variety of
    compositions.  Red dashed lines represent $20\%$ relative error of prediction. 
    Symbols have the same
    meaning as in Figure~\ref{fig:HSB_scalings_comp}.
  }
  \label{fig:HSB_comparison_comp}
\end{figure}

The data collapse of Figure~\ref{fig:HSB_DSi_vs_sx_comp} suggests that 
tracer diffusivities, $D_1$ and $D_2$, of this mixture might also be
adequately predicted using Eq.~\eqref{eq:Dgri_fit} together with
knowledge of the pair potentials, composition, 
and excess entropy of the mixture. Specifically, the generalized Rosenfeld
scaling prediction for tracer diffusivity of component $i$ is given by
\begin{equation}
  \label{eq:Dpredicted_sx}
  D_i^{\RM{predicted}}(\sx)=\left\{\rhoDoi \left[B+T\frac{d B}{ d
        T}\right]\right\} D^{\RM{GR}}_{\RM{SC}}(\sx)
\end{equation}
with  $B$ from Eq.~\eqref{eq:B2}, 
$\rhoDoi$ from Eq.~\eqref{eq:rhoD0_enskog_HS}, and 
$D^{\RM{GR}}_{\RM{SC}}(\sx)$ from the fit of the single-component
data to Eq.~\eqref{eq:Dgri_fit}.

One way to quantitatively assess the relative predictive ability of 
Eq.~\eqref{eq:Dpredicted_sx} is to compare it to
the results of, e.g., the basic Enskog theory given by 
Eq.~\eqref{eq:rhoD_enskog_HS}.  Both equations require as inputs
several pieces of information, including 
the form of the pair potentials and the 
mixture composition.  While predictions based on the 
generalized Rosenfeld scaling
also require knowledge of $\sx$ for the mixture and properties of 
the single-component system, Enskog theory requires
knowledge of the state dependent contact values of the three partial radial
distribution functions of the mixture.  

Figure~\ref{fig:HSB_comparison_comp}
displays the relative error in the predicted tracer diffusivity to the
simulated diffusivity $D_{i}^{\RM{predicted}}/D_{i}$. Enskog theory
[Figure~\ref{fig:HSB_comparison_comp_enskog}] provides good
predictions at high values of tracer diffusivity (i.e., low $\Pc$).  
However, as the value of $D_i$ decreases (i.e. , $\Pc$ increases) 
Enskog theory first underpredicts, then ultimately significantly
overpredicts $D_i$.  When looking at the entire range of $\Pc$ studied
here, $80\%$ of the
Enskog theory predictions lie within
$20\%$ of the molecular simulation data.  On the other hand, the
excess entropy based expression of
Eq.~\eqref{eq:Dpredicted_sx}
[Figure~\ref{fig:HSB_comparison_comp_sx}]
predicts the tracer diffusivities semi-quantitatively 
for all state points
investigated here ($100\%$ of predictions within $20\%$ of the
simulation data).

\subsubsection{Particle-size asymmetry effects}
\label{sec:hs-size-ratio}

\begin{figure*}
  \centering 

  \mbox {
    \includegraphics[clip]{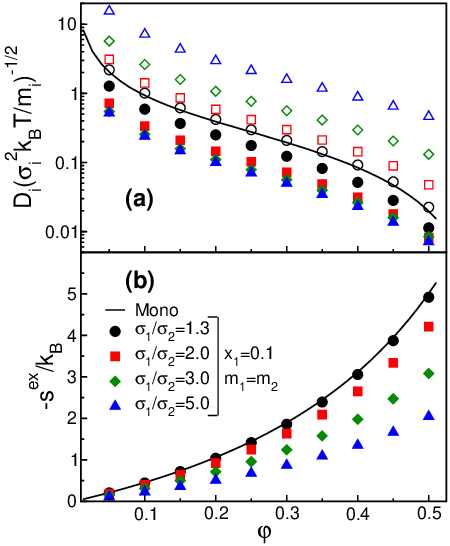}
    \includegraphics[clip]{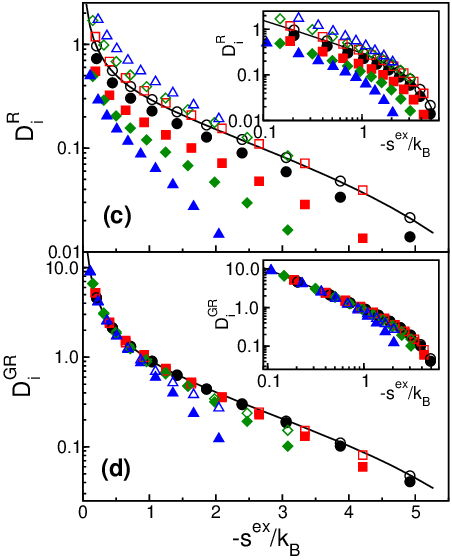}
  }
  
  \subfloat{\label{fig:HSB_D_vs_phic_size_mass}}
  \subfloat{\label{fig:HSB_sx_vs_phic_size_mass}}
  \subfloat{\label{fig:HSB_DRi_sx_size_mass}}
  \subfloat{\label{fig:HSB_DSi_vs_sx_size_mass}}
  \caption {(Color online) Properties of the binary HS mixture at
    composition $x_1=0.1$, equal mass, and several size ratios
    $(\sigma_1/\sigma_2)$ [see legend in (a)]. (a) Tracer diffusion
    coefficient $D_i$ and (b) (negative) excess entropy $-\sx$ versus packing fraction $\Pc$.
    (c) Tracer diffusivity reduced in the original Rosenfeld $\Dri$ and (d) generalized
    Rosenfeld $\Dgri$ forms as a function of $-\sx$.
    Filled and open symbols represent large (component 1) and small
    (component 2) spheres.  Symbol shapes denote different particle
    diameter ratios, as described in the
    legend of (a). The solid line in each figure is the result for the
    single-component HS system.  In (c) and (d) the insets are the same
    as the main plots, but on a log-log scale. }
  \label{fig:HSB_scalings_size_mass}
\end{figure*}

We also study the effects of particle-size asymmetry on the relationship
between excess entropy and tracer diffusivity by examining a series of
binary HS mixtures at composition $x_1=0.1$ and packing fractions in
the range $0 < \Pc \le 0.5$.  Particles of types 1 and
2 were taken to have identical masses, but we investigated several systems
with different diameter ratios 
[$\sigma_1/\sigma_2=1.3$, $2.0$, $3.0$ and $5.0$].
Figure~\ref{fig:HSB_D_vs_phic_size_mass} displays the tracer diffusivities,
$D_1$ and $D_2$, for these systems.  Increasing the magnitude of 
the diameter ratio leads to progressively larger 
deviation of the tracer diffusivities from the
self-diffusion coefficient~$D$ of the HS fluid at the same $\Pc$. As
expected, larger particles diffuse slower than
smaller particles [$D_1<D<D_2$].  How is the excess entropy
affected by increasing the ratio of particle diameters?
Figure~\ref{fig:HSB_sx_vs_phic_size_mass} shows that increasing 
$\sigma_1/\sigma_2$ at fixed $\Pc$ and $x_1$ systematically 
decreases $-\sx$ (i.e.,
weakens the static interparticle correlations).  This
effect is qualitatively connected to the more efficient packing
arrangements that spheres can sample when significant polydispersity is
present \cite{Biazzo2009Theory-of-Amorp,Yerazunis1965Dense-Random-Pa}.

Given that increasing size ratio uniformly reduces structural correlations,
but impacts the dynamics of large and small particles in different
ways, it might not be surprising that tracer diffusivity data represented
in the original Rosenfeld form, $\Dri$, does not collapse when plotted versus
excess entropy [Fig.~\ref{fig:HSB_DRi_sx_size_mass}].  
The $\Dri$ data for the smallest size ratio $\sigma_1/\sigma_2=1.3$ is
qualitatively similar to the single-component result.  However, there
is significant deviation for larger diameter ratios, with excess
entropy underpredicting the mobility of smaller particles and
overpredicting that of larger particles.  
Figure~\ref{fig:HSB_DSi_vs_sx_size_mass} shows, however, that tracer
diffusivity reduced in the generalized Rosenfeld form, $\Dgri$, mostly
collapses to the single-component curve 
when plotted versus $-\sx$.  The most pronounced deviations
are for the largest size ratio ($\sigma_1/\sigma_2=3$, $5$) at the highest
packing fractions ($\phi \ge 0.45$).

\begin{figure}[h]
  \centering
  \includegraphics[clip]{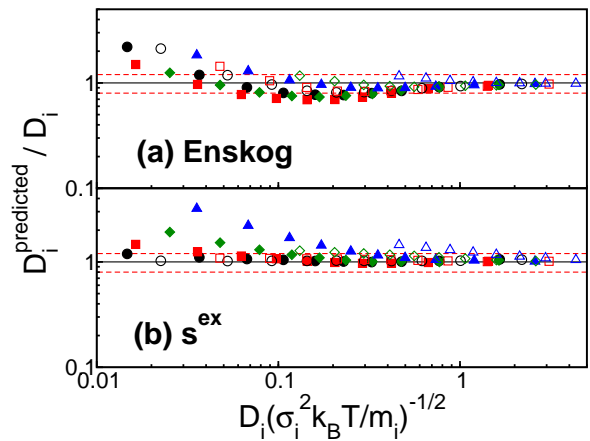}
  \subfloat{\label{fig:HSB_comparison_size_mass_enskog}}
  \subfloat{\label{fig:HSB_comparison_size_mass_sx}}
  \caption{(color online) Ratio of tracer diffusivity predicted from (a)
    Enskog theory [Eq.~\eqref{eq:rhoD_enskog_HS}] and (b) from excess
    entropy and the single-component HS data (i.e., generalized
    Rosenfeld scaling) 
[Eq.~\eqref{eq:Dpredicted_sx}], for a HS mixture with equal
    mass, composition $x_1=0.1$, and a variety of size ratios
    $(\sigma_1/\sigma_2)$.  Red dashed lines represent $20\%$ relative error of prediction.  Symbols have
    the same meaning as in Figure~\ref{fig:HSB_scalings_size_mass}.}
  \label{fig:HSB_comparison_size_mass}
\end{figure}

Figure~\ref{fig:HSB_comparison_size_mass} quantitatively 
compares the predictions of
Enskog theory [Eq.~\eqref{eq:rhoD_enskog_HS}] with those based on
the generalized Rosenfeld scaling [Eq.~\eqref{eq:Dpredicted_sx}].  
At high values of
$D_i$ (low $\Pc$), both methods provide accurate predictions.  As before, for
decreasing $D_i$ (increasing $\Pc$),
Enskog theory first underpredicts and then ultimately overpredicts 
the tracer diffusivities.  The excess entropy based predictions never 
underpredict, but eventually overpredict the mobility at high 
values of $\Pc$.  As an overall measure, the Enskog
and the excess entropy expressions predict $70\%$ and $80\%$ of the
tracer diffusivities within $20\%$ of the simulated values, respectively.
Moreover, we note that while the excess entropy method predicts
the tracer diffusivities of the two components with similar reliability, the
Enskog expression does well for the small particles ($90\%$ within
$20\%$), but poorly for the large particles ($50\%$ within $20\%$). 

\subsubsection{Two-body excess entropy scaling}
\label{sec:dzug-hs}

As noted in Section~\ref{sec:Introduction}, an alternative excess entropy
based scaling for diffusion was introduced by
\citet{Dzugutov1996A-univeral-scal}, who found
that an appropriately reduced form of the self-diffusion coefficient $\Dz$ for
atomic fluids at moderate densities is nearly a universal function of
the two-body contribution to the excess entropy $\stwo$.
Subsequently, others have suggested a generalization of this scaling
\cite{Samanta2001Universal-Scali} to predict tracer diffusivities of
fluid mixtures.  In the generalization, the
reduced tracer-diffusion coefficient, defined as $\Dzi\equiv D_i/\chi_i$, where
\begin{equation}
  \label{eq:chi_dzug}
  \chi_i\equiv4(\pi\kB T)^{1/2} \sum_{j=1}^{N_\RM{c}} x_i \rho
  \sigma_{ij}^4 g_{ij}(\sigma_{ij}^+)\left( \frac{m_i+m_j}{2m_i m_j} \right)^{1/2},
\end{equation}
is thought to approximately scale with the $i$-component contribution
to the two-body
excess entropy, defined as~\cite{Samanta2001Universal-Scali}
\begin{align}
  \label{eq:s2_i}
  \stwoi/\kB&\equiv -\frac{1}{2} \rho \sum_{j=1}^{N_{\RM{c}}} x_j
  \int d\br \nonumber \\
  & \qquad \{g_{ij}(\br) \ln g_{ij}(\br) -[g_{ij}(\br)-1]\}.
\end{align}
Note that the two-body excess entropy per particle is given by
$\stwo=\sum_{i} x_i \stwoi$ \cite{Hernando1990Thermodynamic-p}.
\begin{figure}
  \centering
  \includegraphics[clip]{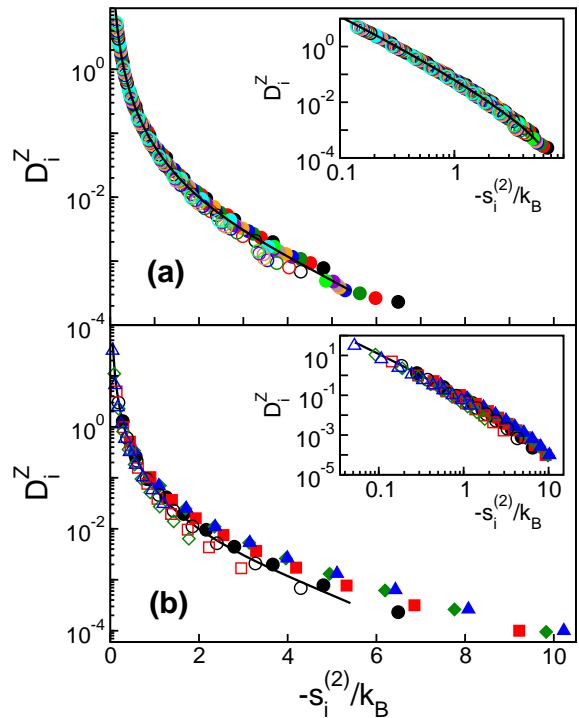}
  \subfloat{\label{fig:HSB_Dzi_s2i_comp}}
  \subfloat{\label{fig:HSB_Dzi_s2i_size_mass}}
    \caption{(color online) Tracer diffusivity reduced in generalized
      Dzugutov form $\Dzi$ discussed in text versus (negative) $i$-component
      contribution to two-body excess entropy $-\stwoi$ of binary HS
      mixtures.  (a) Particle diameter ratio
  $\sigma_1/\sigma_2=1.3$, equal mass, and a variety of compositions. Symbols
  have same meaning as Figure~\ref{fig:HSB_scalings_comp}. (b)
  Composition $x_1=0.1$, equal mass, and a variety of size ratios.  Symbols
  have same meaning as Figure~\ref{fig:HSB_scalings_size_mass}.
  Insets are the same as the main plots, but with a log-log scale.}
  \label{fig:HSB_Dzug}
\end{figure}

In Figure~\ref{fig:HSB_Dzi_s2i_comp} and
\subref*{fig:HSB_Dzi_s2i_size_mass} we examine 
the mixture generalization of the 
Dzugutov scaling for the HS systems discussed above.  Specifically,
we show data with fixed diameter ratio and varying composition in
panel (a) and fixed composition and varying diameter ratio in panel (b). 
Both sets of data more or less track the scaling.  However, deviations
from the single component curve appear systematic. 
The single-component relation with $\stwoi$ overpredict the small 
sphere mobility and underpredict the large sphere mobility. 

Unfortunately, the predictive value of this type of 
scaling is inherently limited by the fact that the
single-component data cannot access the large values of $-\stwoi$
realized by the large spheres in a mixture. The former reach a value 
of $5.5\kB$ at $\Pc=0.5$, while the latter are greater than $10\kB$
for the largest size ratios examined here. As a result, ``predicting''
tracer diffusivities of a mixture would require some systematic way of
extrapolating the single-component curve by a substantial amount.  
As discussed in Section~\ref{sec:Introduction}, the scaling is 
also limited to systems, like the HS fluid, for which
the interparticle repulsions are steep enough to define an effective
hard-core diameter to each interaction.  Thus, it will be of little
use for studying systems with bounded interactions like the
Gaussian-core potential or a other models 
that characterize the soft effective interactions between macro- or
supramolecular species in solution \cite{Likos2001Effective-inter}.

\subsection{Widom-Rowlinson mixtures}
\label{sec:WR}

\begin{figure*}[t]
  \centering
  \mbox{
    \includegraphics[clip]{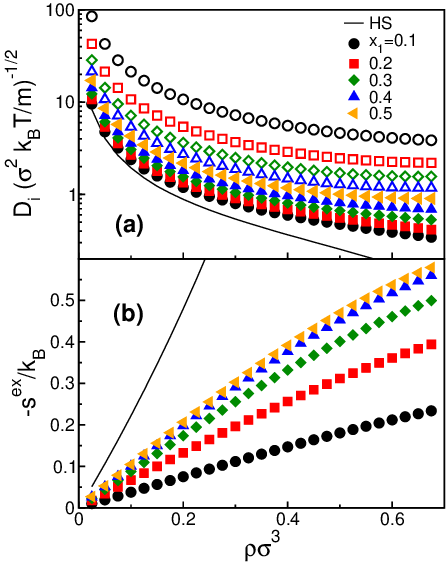}
    \includegraphics[clip]{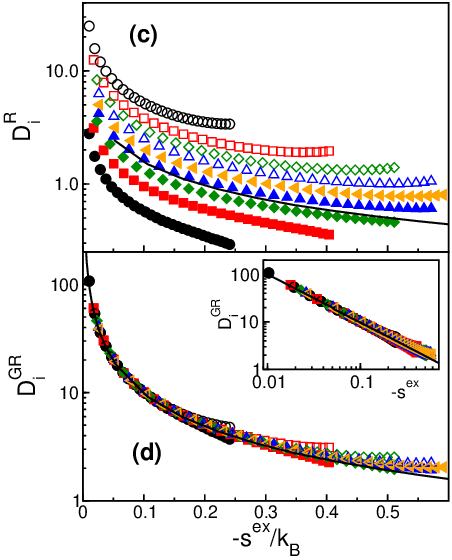}
    
  }
  \subfloat{\label{fig:WR_diff_vs_dens}}
  \subfloat{\label{fig:WR_sx_vs_dens}}
  \subfloat{\label{fig:WR_Dri_sx}}
  \subfloat{\label{fig:WR_Dsi_vs_sx}}
  \caption{(color online)  Properties of the Widom-Rowlinson mixture.
    (a) Tracer diffusivities $D_i$ ($i=1$, $2$) and
    (b) (negative) excess entropy $-\sx$ as function of density $\rho \sigma^3$.  (c)
    Tracer diffusivity reduced in original Rosenfeld form $\Dri$ and
    (d) generalized Rosenfeld form $\Dgri$ as a function of $-\sx$.  
Filled and open
    symbols denote component $1$ and $2$, respectively.  The symbol
    type denotes the mole fraction of component 1, $x_1$, indicated in the
    legend of (a).  The solid line in each figure is the result for
    the single-component HS fluid.  In (d) the insets are the same as the
    main plots, but on a log-log scale. }
  \label{fig:WR_scalings}
\end{figure*}

Here we examine the behavior of the Widom-Rowlinson (WR) model fluid
introduced in Section~\ref{sec:methods}.  Recall that it is defined as a
mixture of non-additive hard spheres with $\sigma_{11}=\sigma_{22}=0$ but
$\sigma_{12}=\sigma$.  Figure~\ref{fig:WR_diff_vs_dens} displays the
tracer diffusivity $D_i$ as a function of density for a several
compositions $x_1$.  Note that the $D_i$ is always greater then the
self-diffusion of the single-component HS fluid, since the number of
collisions per unit time will clearly be less in the 
WR fluid than in the HS fluid at
the same density.  At $x_1=0.5$, $D_1=D_2$,
since the fluid is symmetric.  At fixed density, as $x_1$ decreases, $D_1$ decreases
while $D_2$ increases.  This is because the dilute species 
will experience many more collisions per unit time 
(it has more neighbors of the opposite type) than the concentrated species.
Likewise, Figure~\ref{fig:WR_sx_vs_dens} shows $-\sx$ for the WR
fluids is always less than that of a
single-component HS fluid of the same density.  This is expected since
particles of the same type do not directly exclude volume from one
another, which in turn reduces the entropic driving force for forming
strong interparticle correlations.   Decreasing $x_1$ from 0.5 toward
zero at fixed density decreases
$-\sx$ because it increases the number of particles in the system 
that do not interact.  

As was the case for the HS fluid mixtures, Figure~\ref{fig:WR_Dri_sx}
shows that the tracer diffusivity reduced in the original Rosenfeld
form, $\Dri$, is not even approximately a single-valued function of $\sx$.
On the other hand, Figure~\ref{fig:WR_Dsi_vs_sx} shows that the tracer
diffusivity data
cast in the generalized Rosenfeld form, $\Dgri$, largely collapses
when plotted versus excess entropy.  Interestingly, $\Dgri$ of the WR
fluid is well described by the mathematical form of the 
single-component HS data.  The quality of the
collapse is more easily seen on a log-log scale (inset to
Figure~\ref{fig:WR_Dsi_vs_sx}).

\begin{figure}[h]
  \centering
  \includegraphics[clip]{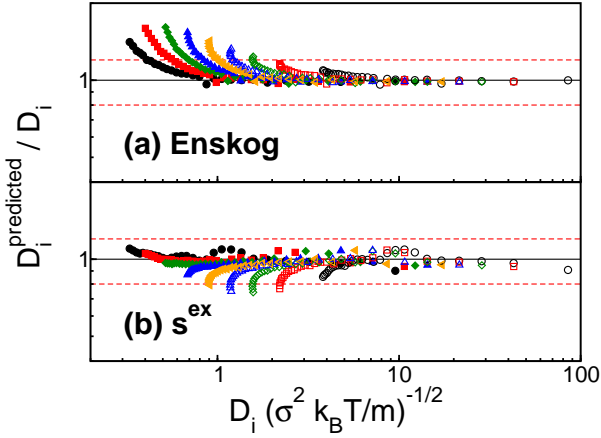}
  \subfloat{\label{fig:WR_comparison_enskog}}
  \subfloat{\label{fig:WR_comparison_sx}}
  \caption{(color online) Ratio of tracer diffusivity predicted from (a)
    Enskog theory [Eq.~\eqref{eq:rhoD_enskog_HS}] and (b) from excess
    entropy and the single-component HS result (generalized Rosenfeld scaling)
    [Eq.~\eqref{eq:Dpredicted_sx}], for the Widom-Rowlinson mixture.
    Red dashed lines indicate $20\%$ relative error of
    prediction.  Symbols have the same meaning as in
    Figure~\ref{fig:WR_scalings}. }
  \label{fig:WR_comparison}
\end{figure}

Figure~\ref{fig:WR_comparison_enskog} and
\subref*{fig:WR_comparison_sx} compare the accuracy of predicting
tracer diffusivity of the components of the WR fluid based on Enskog theory
[Eq.~\eqref{eq:rhoD_enskog_HS}] versus excess entropy of the WR mixture and the
single-component relation for the HS fluid 
[Eq.~\eqref{eq:Dpredicted_sx}].  As for the HS fluid, Enskog theory
predicts $80\%$ of the WR data within $20\%$ of the simulation values.  
In contrast, the excess
entropy method predicts $97\%$ of the data within $20\%$ of the
simulated tracer diffusivities.  

\begin{figure}[h]
  \centering
  \includegraphics[clip]{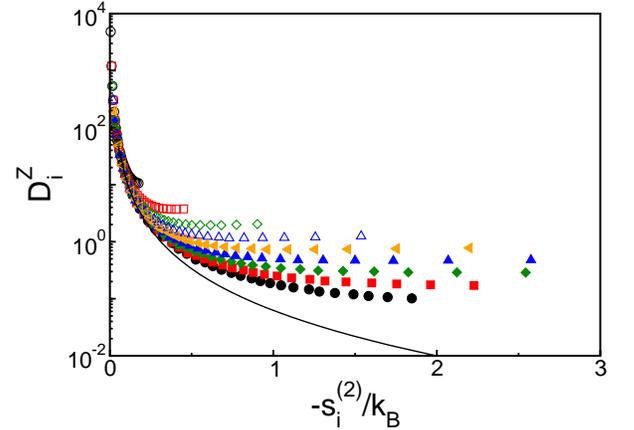}
  \caption{(color online)  Tracer diffusivity reduced in the mixture 
generalized Dzugutov form discussed in text $\Dzi$
  versus the (negative) $i$-component of the two-body excess entropy $\stwoi$ for the 
Widom-Rowlinson model.  Symbols
  have same the meaning as Figure~\ref{fig:WR_scalings}.}
  \label{fig:WR_Dzug}
\end{figure}

Lastly, since the WR model is composed of (non-additive) hard particles, 
it represents another good test case for the mixture generalization of the 
Dzugutov scaling.
Figure~\ref{fig:WR_Dzug}, however,  
clearly shows that this two-body scaling does not collapse the
WR data.  In general, particles of type $i$ 
diffuse considerably faster than would
be predicted based on the single-component HS fluid behavior
and the $i$-component of the two-body excess entropy in the WR
mixture.  Moreover, the magnitude of the under prediction depends
sensitively on composition.
This breakdown of the mixture version of the Dzugutov relation for
non-additive HS fluids indicates that it is not as widely applicable, even
within the limited class of HS model fluids, as the generalized
Rosenfeld scaling introduced here.    

\subsection{Single-component Gaussian-core fluid}
\label{sec:GC-mono}

\begin{figure*}[t]
  \centering
  \mbox{
    \includegraphics[clip]{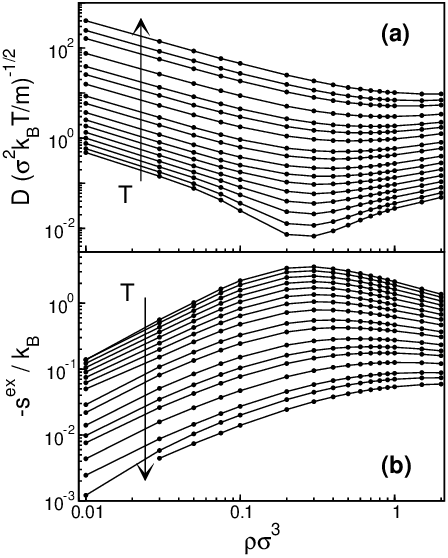}
    \includegraphics[clip]{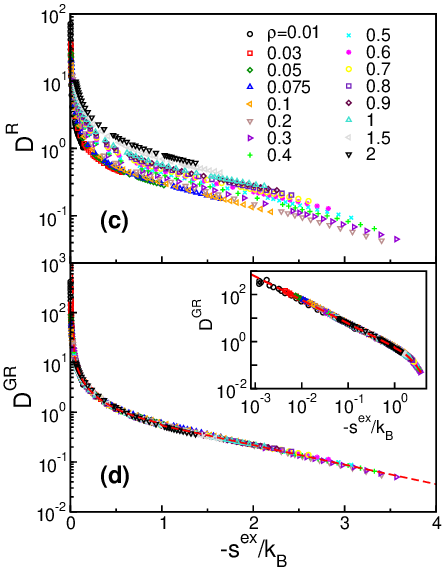}
  }

  \subfloat{\label{fig:GCM_D_vs_rho}}
  \subfloat{\label{fig:GCM_sx_vs_rho}}
  \subfloat{\label{fig:GCM_Dr_vs_sx}}
  \subfloat{\label{fig:GCM_Ds_vs_sx}}
  \caption{(Color online) Properties of the single-component
  Gaussian-core fluid. (a) Self diffusivity
    $D$ and (b) (negative) excess entropy $-\sx$ versus density $\rho$.
    (c) Self diffusivity reduced in the Rosenfeld form $\Dr$ and (d)
    the generalized
    Rosenfeld form $\Dgr$ as a function of $-\sx$.
    In (a) and (b), arrows indicate increasing temperature $T$.  
In (c) and (d), symbol type corresponds to density,
    indicated in the legend of (c).  In (d), inset is the same as the
    main plot but on log-log scale. In (d), the red dashed line
    represents a least-square fit of the data to Eq.~\eqref{eq:Dgri_fit}. 
}
  \label{fig:GCMono}
\end{figure*}

As discussed in Section~\ref{sec:Introduction}, and much more
extensively in
\cite{Mausbach2006Static-and-dyna,Wensink2008Long-time-self-,Krekelberg2009Anomalous-struc,Pond2009Composition-and},
the properties of the single-component Gaussian-core fluid are
anomalous compared to those of simple atomic liquids.  For example, as
shown in Figure~\ref{fig:GCM_D_vs_rho}, the self diffusivity $D$ of
the Gaussian-core fluid first decreases, and then anomalously increases as a
function of particle density along an isotherm.  Likewise, $-\sx$ at
constant temperature first increases (the fluid becomes more
structured), and then anomalously decreases (the fluid becomes less
structured) as a function of density.

In brief, these unusual trends can be qualitatively rationalized based
on the Gaussian form of the repulsion. When the density and
temperature are sufficiently low, the distance between particles is
larger than the range of the potential.  Under these conditions, the
part of the interaction that the particles sample when they
``collide'' appears steeply repulsive, and thus the effects of density
on structure and dynamics are similar to those of HS fluid.  However,
at high particle densities, particles in the Gaussian-core fluid effectively
overlap one another due to the bounded form of the interaction. The
effect is that each particle constantly experiences largely canceling
soft repulsive forces of many neighbors.  Increasing the particle
density under these conditions enhances this effect, paradoxically
weakening the structural correlations and increasing the self
diffusivity of the fluid.

It is clear from Figure~\ref{fig:GCM_D_vs_rho} and
\subref*{fig:GCM_sx_vs_rho} that both $D$ and $\sx$ are strongly
correlated for the Gaussian-core fluid.  In Figure~\ref{fig:GCM_Dr_vs_sx}, we show the
self-diffusion coefficient expressed in the original Rosenfeld form,
$\Dr$, as a function of $-\sx$.  As
noted previously \cite{Krekelberg2009Anomalous-struc}, this basic
scaling is not even approximately a single-valued function of excess
entropy.  However, similar to the behavior of the HS and WR mixtures
discussed previously, Figure~\ref{fig:GCM_Ds_vs_sx}
shows that the generalized-Rosenfeld-scaled self diffusivity $\Dgri$ 
collapses to a single curve when plotted versus excess
entropy.  The quality of the collapse at even low density is
apparent in the log-log plot shown in the inset to
Figure~\ref{fig:GCM_Ds_vs_sx}.  We also observe that the same
functional form that was used to fit the single-component 
HS data can also be applied
to the Gaussian-core system.  In particular, fitting the data to
Eq.~\eqref{eq:Dgri_fit} yields $\alpha=0.59$, $A=1.33$, and $B=0.90$.
As shown in Figure~\ref{fig:GCM_Ds_vs_sx}, Eq.~\eqref{eq:Dgri_fit}
with these parameters (red dashed line) describes the simulation 
data very well.

\subsection{Binary Gaussian-core mixtures}
\label{sec:GC-binary}

\begin{figure*}
  \centering
  \mbox{
    \includegraphics[clip]{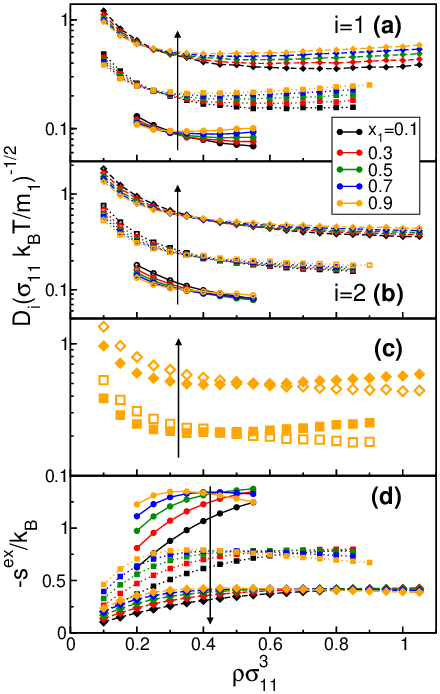}
    \includegraphics[clip]{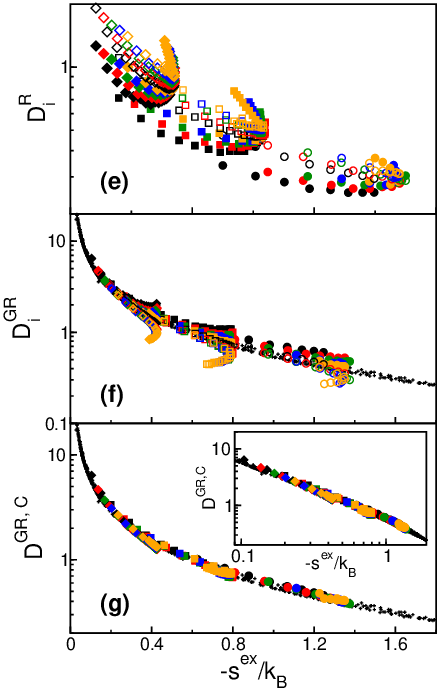}
  }
  \subfloat{\label{fig:GCB_D_vs_rho_split_1}}
  \subfloat{\label{fig:GCB_D_vs_rho_split_2}}
  \subfloat{\label{fig:GCB_D_vs_rho_cross}}
  \subfloat{\label{fig:GCB_D_vs_sx}}
  \subfloat{\label{fig:GCB_DRi_vs_sx}}
  \subfloat{\label{fig:GCB_DSi_vs_sx}}
  \subfloat{\label{fig:GCB_DSC_vs_sx}}
  \caption{(color online) Properties of the binary Gaussian-core
    fluid described in the text.  Tracer-diffusion coefficient of (a)
    component $1$ and (b) component $2$ versus density.  (c)
    Illustration of the crossover in tracer diffusivity for components
    $1$ and $2$ as a function of density. (d) Excess-entropy $-\sx$
    versus density. (e) Rosenfeld scaled tracer-diffusivity $\Dri$,
    (f) generalized Rosenfeld tracer diffusivity $\Dgri$, and
    ``collective'' generalized Rosenfeld tracer diffusivity
    $\DgrC=(\Dgri[1])^{x_1}(\Dgri[2])^{x_2}$ versus $-\sx$.  Symbol
    type corresponds to reduced temperature $\kB T/\epsilon_{11}$:
    $0.05$ (circles), $0.1$ (squares), and $0.2$ (diamonds). For
    clarity in (a)-(d), increasing temperature is given by the
    direction of the arrow..  In (a)-(f), closed and
    open symbols denote components $1$ and $2$, respectively.  Panel
    (c) displays the crossover behavior of the tracer diffusivities
    for $x_1=0.9$ and $\kB T/\epsilon_{11}=0.05$ and $0.1$. In panels (f) and (g), small
    black crosses represent the single-component Gaussian-core data.  Inset to
    (g) is the same as the main plot but on a log-log scale.}
  \label{fig:GCBinary}
\end{figure*}

As a final test of the relationship between single-particle dynamics
and excess entropy in soft-particle fluids, we examine the binary
mixture of Gaussian-core particles described in Section~\ref{sec:methods}.  In
particular,
Figures~\ref{fig:GCB_D_vs_rho_split_1}--\subref*{fig:GCB_D_vs_rho_split_2}
display the tracer diffusivities of the large and small Gaussian-core particles
as a function of density for a variety of mixture compositions
($x_1=0.1$, $0.3$, $0.5$, $0.7$, and $0.9$) and reduced temperatures
$\kB T/ \epsilon=0.05$, $0.1$, and $0.2$.  The first point of interest
in the data, evident in Figure~\ref{fig:GCB_D_vs_rho_split_1}, is that
the tracer diffusivity of the larger type $1$ particles displays the
same anomalous trend as a function of density as the single component
Gaussian-core fluid.  That is, increasing the density eventually leads to an
anomalous increase in $D_1$.  However, over the density range
considered here, the tracer diffusivity of the small type $2$
particles does not show this anomalous trend (see
Figure~\ref{fig:GCB_D_vs_rho_split_2}).  From a qualitative
perspective, these different behaviors perhaps might be expected,
since the larger particles begin to overlap more (and hence transition into
anomalous, mean-field behavior) at lower densities than the smaller
particles.  This aspect of binary Gaussian-core mixtures in discussed
in detail elsewhere \cite{Pond2009Composition-and}.

One consequence of the dynamic decoupling of small and large particles
described above is a crossover density for tracer diffusivity.
Specifically, large particles have lower tracer diffusivity than small
particles at low density, but they attain higher values of tracer
diffusivity than small particles at sufficiently high density [see
Figure~\ref{fig:GCB_D_vs_rho_cross}].  Because of this crossover, the
Gaussian-core fluid mixture is an interesting counterexample to the fluids
discussed thus far.  It appears that both components cannot scale in a
simple way with a single static measure like $-\sx$
[Figure~\ref{fig:GCB_D_vs_sx}].  This is evident, when one considers
how the reduced Rosenfeld [Figure~\ref{fig:GCB_D_vs_sx}] and
generalized Rosenfeld [Figure~\ref{fig:GCB_DRi_vs_sx}] forms of tracer
diffusivity behave as a function of $-\sx$.  As before, the original
Rosenfeld form, $\Dri$, fails to collapse any of the data.  The
generalized Rosenfeld form, $\Dgri$, does an excellent job of
collapsing the low-density data, but necessarily breaks down at higher
densities, where the anomalous behavior emerges.

We close our discussion of the binary Gaussian-core mixture with an interesting 
empirical observation.  In the spirit of \cite{Hoyt2000Test-of-the-Uni}, we find that a
collective tracer diffusivity of the mixture, which we define here as
$\DgrC\equiv (\Dgri[1])^{x_1}(\Dgri[2])^{x_2}$, is still a single-valued of
excess entropy over the wide range of temperature, density, and 
compositions investigated here [see Figure~\ref{fig:GCB_DSC_vs_sx}].  
As can be seen, it also quantitatively tracks the relationship between $\Dr$
and $-\sx$ for the single-component Gaussian-core fluid.  What this implies is
that tracer diffusivity of one component can be predicted based on
knowledge of tracer diffusivity of the other component, the excess
entropy of the mixture, and the behavior of the one-component fluid.
Of
course, this observation also holds true (trivially) for the other
mixtures we discussed earlier because the generalized Rosenfeld tracer 
diffusivities themselves are single-valued functions of $\sx$ of those 
systems.

\section{Conclusions}
\label{sec:conclusions}

In this work, we present a new dimensionless form of the tracer
diffusion coefficient of a species which we call the generalized Rosenfeld 
tracer diffusivity.  We show, via molecular simulation, that this quantity 
is approximately a single-valued
function of excess entropy for a range of model one- and
two-component fluid mixtures.
The empirical excess entropy scaling is consistent with the various 
effects that 
composition, temperature, density, and
microscopic interactions have on the equilibrium single-particle
dynamics of these systems.
Generalizing an earlier argument of
\citet{Rosenfeld1999A-quasi-univers}, we show that the functional 
form of the reduced tracer diffusivity can be obtained by examining the 
theoretical behavior of excess entropy and tracer diffusivity in the
low-particle-density limit.  

We demonstrate that the aforementioned ``generalized Rosenfeld'' scaling applies 
more broadly than other simple approaches such as 
Enksog theory or empirical 
scalings based on the pair-correlation
contribution to the excess entropy.  However, we also identify some
important limitations of the approach.  For example, the scaling
breaks down for highly asymmetric hard-sphere mixtures (diameter
ratios of 5 or larger) for packing fractions near the freezing transition.
It also breaks down for Gaussian-core mixtures, where the softness 
of the interactions combined with the size asymmetry gives rise to
significant decoupling of the single-particle dynamics of the species.  
Interestingly, even in this latter case, we show that a single, 
collective measure
of the tracer diffusivities obeys an excess entropy scaling, which
provides a quantitative link between structure and 
the tracer diffusivities of the two components.

It may also be fruitful in future work to develop generalized
Rosenfeld scalings for other transport coefficients, like thermal
conductivity and shear viscosity.  We plan to focus 
on extending the ideas of the present study 
to systems with other types of dynamics
(e.g., including effects of dissipation, hydrodynamic interactions, etc.).
We also are studying what aspects of interparticle interactions can
give rise to decoupling of species-specific structural and dynamic quantities.

{\bf Acknowledgments}
Two authors (T.M.T and J.R.E) acknowledge financial
support of the National Science Foundation (CTS-
0448721 and CTS-028772, respectively). One author
T.M.T. also acknowledges support of the Welch Foundation
(F-1696) and the David and Lucile Packard Foundation.
The Texas Advanced Computing Center
(TACC) and the Biowulf Cluster at the National Institutes of Health
provided computational resources for this study.

\appendix*

\section{Extension to Brownian dynamics}
\label{sec:supplementary}

Consider a collection of Brownian particles of radius $a$ and volume
fraction $\Pc=4 \pi a^3 \rho/3$ suspended in a continuum solvent. 
We wish to identify a generalized Rosenfeld scaling of $D$, the long-time
self-diffusion coefficient of the particles, which will be 
(approximately) a single-valued 
function of the excess entropy.  In this case, since
the solvent can be being treated as a continuum, the excess entropy of
interest is that associated with the static correlations of the
Brownian particles.  

At infinite dilution, the excess entropy of the Brownian particles is
zero,
and $D$ is simply equal to the Stokes-Einstein diffusivity
$D_0=\kB T/(S\pi\eta a)$ where $S=4$ for slip and $S=6$ for stick
boundary conditions, and $\eta$ is the solvent viscosity.  To leading
order in packing fraction $\Pc$, the difference between $D_0$ and $D$
can be expressed
\cite{Cichocki1990Diffusion-coeff,Cichocki1988Long-time-self-,Batchelor1983Diffusion-in-a-,Batchelor1976Brownian-Diffus}
\begin{equation}
  \label{eq:D_hydrodynamic}
  \Delta D\equiv D_0-D=- D_0 D_2 \Pc,
\end{equation}
where $D_2$ characterizes how
static correlations modify the long-time self-diffusivity of the particles.
This quantity can be expressed as
\begin{align}
  D_2&=\int_0^{\infty}(-3 +A_{11}+2 B_{11})g(r)r^2 dr + \nonumber \\
  &\quad \quad \int_0^{\infty} \bigg[ \frac{A_{11}-A_{12}-B_{11}+B_{12}}{r} + \nonumber \\
  &\quad \qquad \frac{1}{2} \left( \frac{d A_{11}}{d r}-\frac{d A_{12}}{d r}\right) \bigg] Q(r) g(r) r^2 dr. \label{eq:D2}
\end{align}
Expressions for the functions $Q(r)$, $A_{11}$, $A_{12}$, $B_{11}$,
and $B_{12}$ are known in the hydrodynamic limit
\cite{Jeffrey1984Calculation-of-,Cichocki1988Long-time-self-}.  Also,
in the dilute limit, one may further replace $g(r)$ in
Eq.~\eqref{eq:D2} by the Boltzmann factor
[Eq.~\eqref{eq:pcf_low_dens}] of  $V'(r)$, the effective pair potential between
the Brownian particles in solution. 

Following Eq.~\eqref{eq:sx_second_virial}, one can similarly express
the excess entropy 
$s'^{\RM{ex}}$ associated with the structure of the Brownian particles
(to leading order in particle density) in terms of its osmotic second 
virial coefficient, $s'^{\RM{ex}}=-\kB
\rho [B'+ dB'/d \ln T]$.  As in Eq.~\eqref{eq:Bij}, $B'$ can be readily 
obtained from knowledge of $V'(r)$.  Using the osmotic virial
expression for the excess entropy to eliminate packing fraction from 
Eq.~\eqref{eq:D_hydrodynamic} and rearranging yields a
Brownian-appropriate generalized Rosenfeld self diffusivity:
\begin{equation}
  \label{eq:GR_hydrodynamic}
  \Delta D^{\RM{GR}}\equiv\frac{\left( D_0-D \right) \left[B'+T
      \frac{dB'}{dT}\right]}{-D_0 D_2 \frac{4}{3}\pi a^3}= -s'^{\RM{ex}}/\kB.
\end{equation}
Although the quantity, $\Delta D^{\RM{GR}}$ is strictly a
single-valued function of $s'^{\RM{ex}}$ in the dilute limit, its
behavior at higher particle concentrations needs to be studied
further.  


\begin{thebibliography}{65}
\expandafter\ifx\csname natexlab\endcsname\relax\def\natexlab#1{#1}\fi
\expandafter\ifx\csname bibnamefont\endcsname\relax
  \def\bibnamefont#1{#1}\fi
\expandafter\ifx\csname bibfnamefont\endcsname\relax
  \def\bibfnamefont#1{#1}\fi
\expandafter\ifx\csname citenamefont\endcsname\relax
  \def\citenamefont#1{#1}\fi
\expandafter\ifx\csname url\endcsname\relax
  \def\url#1{\texttt{#1}}\fi
\expandafter\ifx\csname urlprefix\endcsname\relax\def\urlprefix{URL }\fi
\providecommand{\bibinfo}[2]{#2}
\providecommand{\eprint}[2][]{\url{#2}}

\bibitem[{\citenamefont{Rosenfeld}(1977)}]{Rosenfeld1977Relation-betwee}
\bibinfo{author}{\bibfnamefont{Y.}~\bibnamefont{Rosenfeld}},
  \bibinfo{journal}{Phys. Rev. A} \textbf{\bibinfo{volume}{15}},
  \bibinfo{pages}{2545} (\bibinfo{year}{1977}).

\bibitem[{\citenamefont{Rosenfeld}(1999)}]{Rosenfeld1999A-quasi-univers}
\bibinfo{author}{\bibfnamefont{Y.}~\bibnamefont{Rosenfeld}},
  \bibinfo{journal}{J. Phys.: Condens. Matter} \textbf{\bibinfo{volume}{11}},
  \bibinfo{pages}{5415} (\bibinfo{year}{1999}).

\bibitem[{\citenamefont{Dzugutov}(1996)}]{Dzugutov1996A-univeral-scal}
\bibinfo{author}{\bibfnamefont{M.}~\bibnamefont{Dzugutov}},
  \bibinfo{journal}{Nature} \textbf{\bibinfo{volume}{381}},
  \bibinfo{pages}{137} (\bibinfo{year}{1996}).

\bibitem[{\citenamefont{Mittal et~al.}(2007{\natexlab{a}})\citenamefont{Mittal,
  Errington, and Truskett}}]{Mittal2007Relationships-b}
\bibinfo{author}{\bibfnamefont{J.}~\bibnamefont{Mittal}},
  \bibinfo{author}{\bibfnamefont{J.}~\bibnamefont{Errington}},
  \bibnamefont{and} \bibinfo{author}{\bibfnamefont{T.}~\bibnamefont{Truskett}},
  \bibinfo{journal}{J. Phys. Chem. B} \textbf{\bibinfo{volume}{111}},
  \bibinfo{pages}{10054} (\bibinfo{year}{2007}{\natexlab{a}}).

\bibitem[{\citenamefont{Abramson}(2007)}]{Abramson2007Viscosity-of-wa}
\bibinfo{author}{\bibfnamefont{E.~H.} \bibnamefont{Abramson}},
  \bibinfo{journal}{Phys. Rev. E} \textbf{\bibinfo{volume}{76}},
  \bibinfo{pages}{051203} (\bibinfo{year}{2007}).

\bibitem[{\citenamefont{Abramson and
  West-Foyle}(2008)}]{Abramson2008Viscosity-of-ni}
\bibinfo{author}{\bibfnamefont{E.~H.} \bibnamefont{Abramson}} \bibnamefont{and}
  \bibinfo{author}{\bibfnamefont{H.}~\bibnamefont{West-Foyle}},
  \bibinfo{journal}{Phys. Rev. E} \textbf{\bibinfo{volume}{77}},
  \bibinfo{pages}{041202} (\bibinfo{year}{2008}).

\bibitem[{\citenamefont{Abramson}(2009)}]{Abramson2009Viscosity-of-ca}
\bibinfo{author}{\bibfnamefont{E.~H.} \bibnamefont{Abramson}},
  \bibinfo{journal}{Phys. Rev. E} \textbf{\bibinfo{volume}{80}},
  \bibinfo{pages}{021201} (\bibinfo{year}{2009}).

\bibitem[{\citenamefont{Mittal et~al.}(2006{\natexlab{a}})\citenamefont{Mittal,
  Errington, and Truskett}}]{Mittal2006Thermodynamics-}
\bibinfo{author}{\bibfnamefont{J.}~\bibnamefont{Mittal}},
  \bibinfo{author}{\bibfnamefont{J.~R.} \bibnamefont{Errington}},
  \bibnamefont{and} \bibinfo{author}{\bibfnamefont{T.~M.}
  \bibnamefont{Truskett}}, \bibinfo{journal}{Phys. Rev. Lett.}
  \textbf{\bibinfo{volume}{96}}, \bibinfo{pages}{177804}
  (\bibinfo{year}{2006}{\natexlab{a}}).

\bibitem[{\citenamefont{Mittal et~al.}(2007{\natexlab{b}})\citenamefont{Mittal,
  Errington, and Truskett}}]{Mittal2007Does-confining-}
\bibinfo{author}{\bibfnamefont{J.}~\bibnamefont{Mittal}},
  \bibinfo{author}{\bibfnamefont{J.~R.} \bibnamefont{Errington}},
  \bibnamefont{and} \bibinfo{author}{\bibfnamefont{T.~M.}
  \bibnamefont{Truskett}}, \bibinfo{journal}{J. Chem. Phys.}
  \textbf{\bibinfo{volume}{126}}, \bibinfo{pages}{244708}
  (\bibinfo{year}{2007}{\natexlab{b}}).

\bibitem[{\citenamefont{Mittal et~al.}(2007{\natexlab{c}})\citenamefont{Mittal,
  Shen, Errington, and Truskett}}]{Mittal2007Confinement-ent}
\bibinfo{author}{\bibfnamefont{J.}~\bibnamefont{Mittal}},
  \bibinfo{author}{\bibfnamefont{V.~K.} \bibnamefont{Shen}},
  \bibinfo{author}{\bibfnamefont{J.~R.} \bibnamefont{Errington}},
  \bibnamefont{and} \bibinfo{author}{\bibfnamefont{T.~M.}
  \bibnamefont{Truskett}}, \bibinfo{journal}{J. Chem. Phys.}
  \textbf{\bibinfo{volume}{127}}, \bibinfo{pages}{154513}
  (\bibinfo{year}{2007}{\natexlab{c}}).

\bibitem[{\citenamefont{Goel et~al.}(2009)\citenamefont{Goel, Krekelberg, Pond,
  Mittal, Shen, Errington, and Truskett}}]{Goel2009Available-state}
\bibinfo{author}{\bibfnamefont{G.}~\bibnamefont{Goel}},
  \bibinfo{author}{\bibfnamefont{W.~P.} \bibnamefont{Krekelberg}},
  \bibinfo{author}{\bibfnamefont{M.~J.} \bibnamefont{Pond}},
  \bibinfo{author}{\bibfnamefont{J.}~\bibnamefont{Mittal}},
  \bibinfo{author}{\bibfnamefont{V.~K.} \bibnamefont{Shen}},
  \bibinfo{author}{\bibfnamefont{J.~R.} \bibnamefont{Errington}},
  \bibnamefont{and} \bibinfo{author}{\bibfnamefont{T.~M.}
  \bibnamefont{Truskett}}, \bibinfo{journal}{J. Stat. Mech.}
  \textbf{\bibinfo{volume}{2009}}, \bibinfo{pages}{P04006}
  (\bibinfo{year}{2009}).

\bibitem[{\citenamefont{Goel et~al.}(2008)\citenamefont{Goel, Krekelberg,
  Errington, and Truskett}}]{Goel2008Tuning-Density-}
\bibinfo{author}{\bibfnamefont{G.}~\bibnamefont{Goel}},
  \bibinfo{author}{\bibfnamefont{W.~P.} \bibnamefont{Krekelberg}},
  \bibinfo{author}{\bibfnamefont{J.~R.} \bibnamefont{Errington}},
  \bibnamefont{and} \bibinfo{author}{\bibfnamefont{T.~M.}
  \bibnamefont{Truskett}}, \bibinfo{journal}{Phys. Rev. Lett.}
  \textbf{\bibinfo{volume}{100}}, \bibinfo{pages}{106001}
  (\bibinfo{year}{2008}).

\bibitem[{\citenamefont{Errington and
  Debenedetti}(2001)}]{Errington2001Relationship-be}
\bibinfo{author}{\bibfnamefont{J.~R.} \bibnamefont{Errington}}
  \bibnamefont{and} \bibinfo{author}{\bibfnamefont{P.~G.}
  \bibnamefont{Debenedetti}}, \bibinfo{journal}{Nature}
  \textbf{\bibinfo{volume}{409}}, \bibinfo{pages}{318} (\bibinfo{year}{2001}).

\bibitem[{\citenamefont{Shell et~al.}(2002)\citenamefont{Shell, Debenedetti,
  and Panagiotopoulos}}]{Shell2002Molecular-struc}
\bibinfo{author}{\bibfnamefont{M.~S.} \bibnamefont{Shell}},
  \bibinfo{author}{\bibfnamefont{P.~G.} \bibnamefont{Debenedetti}},
  \bibnamefont{and} \bibinfo{author}{\bibfnamefont{A.~Z.}
  \bibnamefont{Panagiotopoulos}}, \bibinfo{journal}{Phys. Rev. E}
  \textbf{\bibinfo{volume}{66}}, \bibinfo{pages}{011202}
  (\bibinfo{year}{2002}).

\bibitem[{\citenamefont{Truskett and Dill}(2002)}]{Truskett2002A-Simple-Statis}
\bibinfo{author}{\bibfnamefont{T.~M.} \bibnamefont{Truskett}} \bibnamefont{and}
  \bibinfo{author}{\bibfnamefont{K.~A.} \bibnamefont{Dill}},
  \bibinfo{journal}{J. Phys. Chem. B} \textbf{\bibinfo{volume}{106}},
  \bibinfo{pages}{11829} (\bibinfo{year}{2002}).

\bibitem[{\citenamefont{Kumar et~al.}(2005)\citenamefont{Kumar, Buldyrev,
  Sciortino, Zaccarelli, and Stanley}}]{Kumar2005Static-and-dyna}
\bibinfo{author}{\bibfnamefont{P.}~\bibnamefont{Kumar}},
  \bibinfo{author}{\bibfnamefont{S.~V.} \bibnamefont{Buldyrev}},
  \bibinfo{author}{\bibfnamefont{F.}~\bibnamefont{Sciortino}},
  \bibinfo{author}{\bibfnamefont{E.}~\bibnamefont{Zaccarelli}},
  \bibnamefont{and} \bibinfo{author}{\bibfnamefont{H.~E.}
  \bibnamefont{Stanley}}, \bibinfo{journal}{Phys. Rev. E}
  \textbf{\bibinfo{volume}{72}}, \bibinfo{pages}{021501}
  (\bibinfo{year}{2005}).

\bibitem[{\citenamefont{Esposito et~al.}(2006)\citenamefont{Esposito, Saija,
  Saitta, and Giaquinta}}]{Esposito2006Entropy-based-m}
\bibinfo{author}{\bibfnamefont{R.}~\bibnamefont{Esposito}},
  \bibinfo{author}{\bibfnamefont{F.}~\bibnamefont{Saija}},
  \bibinfo{author}{\bibfnamefont{A.~M.} \bibnamefont{Saitta}},
  \bibnamefont{and} \bibinfo{author}{\bibfnamefont{P.~V.}
  \bibnamefont{Giaquinta}}, \bibinfo{journal}{Phys. Rev. E}
  \textbf{\bibinfo{volume}{73}}, \bibinfo{pages}{040502(R)}
  (\bibinfo{year}{2006}).

\bibitem[{\citenamefont{Netz et~al.}(2006)\citenamefont{Netz, Buldyrev,
  Barbosa, and Stanley}}]{Netz2006Thermodynamic-a}
\bibinfo{author}{\bibfnamefont{P.~A.} \bibnamefont{Netz}},
  \bibinfo{author}{\bibfnamefont{S.~V.} \bibnamefont{Buldyrev}},
  \bibinfo{author}{\bibfnamefont{M.~C.} \bibnamefont{Barbosa}},
  \bibnamefont{and} \bibinfo{author}{\bibfnamefont{H.~E.}
  \bibnamefont{Stanley}}, \bibinfo{journal}{Phys. Rev. E}
  \textbf{\bibinfo{volume}{73}}, \bibinfo{pages}{061504}
  (\bibinfo{year}{2006}).

\bibitem[{\citenamefont{Xu et~al.}(2006)\citenamefont{Xu, Buldyrev, Angell, and
  Stanley}}]{Xu2006Thermodynamics-}
\bibinfo{author}{\bibfnamefont{L.}~\bibnamefont{Xu}},
  \bibinfo{author}{\bibfnamefont{S.~V.} \bibnamefont{Buldyrev}},
  \bibinfo{author}{\bibfnamefont{C.~A.} \bibnamefont{Angell}},
  \bibnamefont{and} \bibinfo{author}{\bibfnamefont{H.~E.}
  \bibnamefont{Stanley}}, \bibinfo{journal}{Phys. Rev. E}
  \textbf{\bibinfo{volume}{74}}, \bibinfo{pages}{031108}
  (\bibinfo{year}{2006}).

\bibitem[{\citenamefont{Mittal et~al.}(2006{\natexlab{b}})\citenamefont{Mittal,
  Errington, and Truskett}}]{Mittal2006Quantitative-Li}
\bibinfo{author}{\bibfnamefont{J.}~\bibnamefont{Mittal}},
  \bibinfo{author}{\bibfnamefont{J.~R.} \bibnamefont{Errington}},
  \bibnamefont{and} \bibinfo{author}{\bibfnamefont{T.~M.}
  \bibnamefont{Truskett}}, \bibinfo{journal}{J. Phys. Chem. B}
  \textbf{\bibinfo{volume}{110}}, \bibinfo{pages}{18147}
  (\bibinfo{year}{2006}{\natexlab{b}}).

\bibitem[{\citenamefont{Errington et~al.}(2006)\citenamefont{Errington,
  Truskett, and Mittal}}]{Errington2006Excess-entropy-}
\bibinfo{author}{\bibfnamefont{J.~R.} \bibnamefont{Errington}},
  \bibinfo{author}{\bibfnamefont{T.~M.} \bibnamefont{Truskett}},
  \bibnamefont{and} \bibinfo{author}{\bibfnamefont{J.}~\bibnamefont{Mittal}},
  \bibinfo{journal}{J. Chem. Phys.} \textbf{\bibinfo{volume}{125}},
  \bibinfo{pages}{244502} (\bibinfo{year}{2006}).

\bibitem[{\citenamefont{Sharma et~al.}(2006)\citenamefont{Sharma, Chakraborty,
  and Chakravarty}}]{Sharma2006Entropy-diffusi}
\bibinfo{author}{\bibfnamefont{R.}~\bibnamefont{Sharma}},
  \bibinfo{author}{\bibfnamefont{S.~N.} \bibnamefont{Chakraborty}},
  \bibnamefont{and}
  \bibinfo{author}{\bibfnamefont{C.}~\bibnamefont{Chakravarty}},
  \bibinfo{journal}{J. Chem. Phys.} \textbf{\bibinfo{volume}{125}},
  \bibinfo{pages}{204501} (\bibinfo{year}{2006}).

\bibitem[{\citenamefont{de~Oliveira et~al.}(2007)\citenamefont{de~Oliveira,
  Barbosa, and Netz}}]{Oliveira2007Interplay-betwe}
\bibinfo{author}{\bibfnamefont{A.~B.} \bibnamefont{de~Oliveira}},
  \bibinfo{author}{\bibfnamefont{M.~C.} \bibnamefont{Barbosa}},
  \bibnamefont{and} \bibinfo{author}{\bibfnamefont{P.~A.} \bibnamefont{Netz}},
  \bibinfo{journal}{Physica A} \textbf{\bibinfo{volume}{386}},
  \bibinfo{pages}{744} (\bibinfo{year}{2007}).

\bibitem[{\citenamefont{Yan et~al.}(2007)\citenamefont{Yan, Buldyrev, Kumar,
  Giovambattista, Debenedetti, and Stanley}}]{Yan2007Structure-of-th}
\bibinfo{author}{\bibfnamefont{Z.}~\bibnamefont{Yan}},
  \bibinfo{author}{\bibfnamefont{S.~V.} \bibnamefont{Buldyrev}},
  \bibinfo{author}{\bibfnamefont{P.}~\bibnamefont{Kumar}},
  \bibinfo{author}{\bibfnamefont{N.}~\bibnamefont{Giovambattista}},
  \bibinfo{author}{\bibfnamefont{P.~G.} \bibnamefont{Debenedetti}},
  \bibnamefont{and} \bibinfo{author}{\bibfnamefont{H.~E.}
  \bibnamefont{Stanley}}, \bibinfo{journal}{Phys. Rev. E}
  \textbf{\bibinfo{volume}{76}}, \bibinfo{pages}{051201}
  (\bibinfo{year}{2007}).

\bibitem[{\citenamefont{Szortyka and
  Barbosa}(2007)}]{Szortyka2007Diffusion-anoma}
\bibinfo{author}{\bibfnamefont{M.~M.} \bibnamefont{Szortyka}} \bibnamefont{and}
  \bibinfo{author}{\bibfnamefont{M.~C.} \bibnamefont{Barbosa}},
  \bibinfo{journal}{Physica A} \textbf{\bibinfo{volume}{380}},
  \bibinfo{pages}{27} (\bibinfo{year}{2007}).

\bibitem[{\citenamefont{de~Oliveira et~al.}(2008)\citenamefont{de~Oliveira,
  Franzese, Netz, and Barbosa}}]{Oliveira2008Waterlike-hiera}
\bibinfo{author}{\bibfnamefont{A.~B.} \bibnamefont{de~Oliveira}},
  \bibinfo{author}{\bibfnamefont{G.}~\bibnamefont{Franzese}},
  \bibinfo{author}{\bibfnamefont{P.~A.} \bibnamefont{Netz}}, \bibnamefont{and}
  \bibinfo{author}{\bibfnamefont{M.~C.} \bibnamefont{Barbosa}},
  \bibinfo{journal}{J. Chem. Phys.} \textbf{\bibinfo{volume}{128}},
  \bibinfo{pages}{064901} (\bibinfo{year}{2008}).

\bibitem[{\citenamefont{Krekelberg et~al.}(2008)\citenamefont{Krekelberg,
  Mittal, Ganesan, and Truskett}}]{Krekelberg2008Structural-anom}
\bibinfo{author}{\bibfnamefont{W.~P.} \bibnamefont{Krekelberg}},
  \bibinfo{author}{\bibfnamefont{J.}~\bibnamefont{Mittal}},
  \bibinfo{author}{\bibfnamefont{V.}~\bibnamefont{Ganesan}}, \bibnamefont{and}
  \bibinfo{author}{\bibfnamefont{T.~M.} \bibnamefont{Truskett}},
  \bibinfo{journal}{Phys. Rev. E} \textbf{\bibinfo{volume}{77}},
  \bibinfo{pages}{041201} (\bibinfo{year}{2008}).

\bibitem[{\citenamefont{Yan et~al.}(2008)\citenamefont{Yan, Buldyrev, and
  Stanley}}]{Yan2008Relation-of-wat}
\bibinfo{author}{\bibfnamefont{Z.}~\bibnamefont{Yan}},
  \bibinfo{author}{\bibfnamefont{S.~V.} \bibnamefont{Buldyrev}},
  \bibnamefont{and} \bibinfo{author}{\bibfnamefont{H.~E.}
  \bibnamefont{Stanley}}, \bibinfo{journal}{Phys. Rev. E}
  \textbf{\bibinfo{volume}{78}}, \bibinfo{pages}{051201}
  (\bibinfo{year}{2008}).

\bibitem[{\citenamefont{Krekelberg et~al.}(2007)\citenamefont{Krekelberg,
  Mittal, Ganesan, and Truskett}}]{Krekelberg2007How-short-range}
\bibinfo{author}{\bibfnamefont{W.~P.} \bibnamefont{Krekelberg}},
  \bibinfo{author}{\bibfnamefont{J.}~\bibnamefont{Mittal}},
  \bibinfo{author}{\bibfnamefont{V.}~\bibnamefont{Ganesan}}, \bibnamefont{and}
  \bibinfo{author}{\bibfnamefont{T.~M.} \bibnamefont{Truskett}},
  \bibinfo{journal}{J. Chem. Phys.} \textbf{\bibinfo{volume}{127}},
  \bibinfo{pages}{044502} (\bibinfo{year}{2007}).

\bibitem[{\citenamefont{Krekelberg et~al.}(2009)\citenamefont{Krekelberg,
  Kumar, Mittal, Errington, and Truskett}}]{Krekelberg2009Anomalous-struc}
\bibinfo{author}{\bibfnamefont{W.~P.} \bibnamefont{Krekelberg}},
  \bibinfo{author}{\bibfnamefont{T.}~\bibnamefont{Kumar}},
  \bibinfo{author}{\bibfnamefont{J.}~\bibnamefont{Mittal}},
  \bibinfo{author}{\bibfnamefont{J.~R.} \bibnamefont{Errington}},
  \bibnamefont{and} \bibinfo{author}{\bibfnamefont{T.~M.}
  \bibnamefont{Truskett}}, \bibinfo{journal}{Phys. Rev. E}
  \textbf{\bibinfo{volume}{79}}, \bibinfo{pages}{031203}
  (\bibinfo{year}{2009}).

\bibitem[{\citenamefont{Pond et~al.}(2009)\citenamefont{Pond, Krekelberg, Shen,
  Errington, and Truskett}}]{Pond2009Composition-and}
\bibinfo{author}{\bibfnamefont{M.~J.} \bibnamefont{Pond}},
  \bibinfo{author}{\bibfnamefont{W.~P.} \bibnamefont{Krekelberg}},
  \bibinfo{author}{\bibfnamefont{V.~K.} \bibnamefont{Shen}},
  \bibinfo{author}{\bibfnamefont{J.~R.} \bibnamefont{Errington}},
  \bibnamefont{and} \bibinfo{author}{\bibfnamefont{T.~M.}
  \bibnamefont{Truskett}}, \bibinfo{journal}{arXiv:0908.3014}
  (\bibinfo{year}{2009}).

\bibitem[{\citenamefont{Chaimovich and
  Shell}(2009)}]{Chaimovich2009Anomalous-water}
\bibinfo{author}{\bibfnamefont{A.}~\bibnamefont{Chaimovich}} \bibnamefont{and}
  \bibinfo{author}{\bibfnamefont{M.~S.} \bibnamefont{Shell}},
  \bibinfo{journal}{Phys. Chem. Chem. Phys.} \textbf{\bibinfo{volume}{11}},
  \bibinfo{pages}{1901} (\bibinfo{year}{2009}).

\bibitem[{\citenamefont{Hoover}(1991)}]{Hoover1991Computational-S}
\bibinfo{author}{\bibfnamefont{W.~G.} \bibnamefont{Hoover}},
  \emph{\bibinfo{title}{Computational Statistical Mechanics}}
  (\bibinfo{publisher}{Elsevier Science Pub Co}, \bibinfo{year}{1991}), pp.
  \bibinfo{pages}{172--173}.

\bibitem[{\citenamefont{Gnan et~al.}(2009)\citenamefont{Gnan, Schrâˆšâˆder,
  Pedersen, Bailey, and Dyre}}]{Gnan2009Pressure-energy}
\bibinfo{author}{\bibfnamefont{N.}~\bibnamefont{Gnan}},
  \bibinfo{author}{\bibfnamefont{T.~B.} \bibnamefont{Schrâˆšâˆder}},
  \bibinfo{author}{\bibfnamefont{U.~R.} \bibnamefont{Pedersen}},
  \bibinfo{author}{\bibfnamefont{N.~P.} \bibnamefont{Bailey}},
  \bibnamefont{and} \bibinfo{author}{\bibfnamefont{J.~C.} \bibnamefont{Dyre}},
  \bibinfo{journal}{arXiv:0905.3497v1}  (\bibinfo{year}{2009}).

\bibitem[{\citenamefont{Lang et~al.}(2000)\citenamefont{Lang, Likos, Watzlawek,
  and Lowen}}]{Lang2000Fluid-and-solid}
\bibinfo{author}{\bibfnamefont{A.}~\bibnamefont{Lang}},
  \bibinfo{author}{\bibfnamefont{C.~N.} \bibnamefont{Likos}},
  \bibinfo{author}{\bibfnamefont{M.}~\bibnamefont{Watzlawek}},
  \bibnamefont{and} \bibinfo{author}{\bibfnamefont{H.}~\bibnamefont{Lowen}},
  \bibinfo{journal}{J. Phys.: Condens. Matter} \textbf{\bibinfo{volume}{12}},
  \bibinfo{pages}{5087} (\bibinfo{year}{2000}).

\bibitem[{\citenamefont{Louis et~al.}(2000)\citenamefont{Louis, Bolhuis, and
  Hansen}}]{Louis2000Mean-field-flui}
\bibinfo{author}{\bibfnamefont{A.~A.} \bibnamefont{Louis}},
  \bibinfo{author}{\bibfnamefont{P.~G.} \bibnamefont{Bolhuis}},
  \bibnamefont{and} \bibinfo{author}{\bibfnamefont{J.~P.}
  \bibnamefont{Hansen}}, \bibinfo{journal}{Phys. Rev. E}
  \textbf{\bibinfo{volume}{62}}, \bibinfo{pages}{7961} (\bibinfo{year}{2000}).

\bibitem[{\citenamefont{Likos}(2001)}]{Likos2001Effective-inter}
\bibinfo{author}{\bibfnamefont{C.~N.} \bibnamefont{Likos}},
  \bibinfo{journal}{Phys. Rep.} \textbf{\bibinfo{volume}{348}},
  \bibinfo{pages}{267} (\bibinfo{year}{2001}).

\bibitem[{\citenamefont{Mausbach and May}(2009)}]{Mausbach2009Transport-}
\bibinfo{author}{\bibfnamefont{P.}~\bibnamefont{Mausbach}} \bibnamefont{and}
  \bibinfo{author}{\bibfnamefont{H.-O.} \bibnamefont{May}},
  \bibinfo{journal}{Z. Phys. Chem..} \textbf{\bibinfo{volume}{223}},
  \bibinfo{pages}{1035} (\bibinfo{year}{2009}).

\bibitem[{\citenamefont{Samanta et~al.}(2001)\citenamefont{Samanta, Ali, and
  Ghosh}}]{Samanta2001Universal-Scali}
\bibinfo{author}{\bibfnamefont{A.}~\bibnamefont{Samanta}},
  \bibinfo{author}{\bibfnamefont{S.~M.} \bibnamefont{Ali}}, \bibnamefont{and}
  \bibinfo{author}{\bibfnamefont{S.~K.} \bibnamefont{Ghosh}},
  \bibinfo{journal}{Phys. Rev. Lett.} \textbf{\bibinfo{volume}{87}},
  \bibinfo{pages}{245901} (\bibinfo{year}{2001}).

\bibitem[{foo()}]{footnote1}
\bibinfo{note}{The focus of the present paper is on molecular dynamics, where
  the particles obey Newton's equations of motion. In the Appendix, we
  illustrate how this analysis might be extended to treat the dissipative
  dynamics relevant for Brownian particles suspended in a solvent.}

\bibitem[{\citenamefont{Chapman and
  Cowling}(1970)}]{Chapman1970The-Mathematica}
\bibinfo{author}{\bibfnamefont{S.}~\bibnamefont{Chapman}} \bibnamefont{and}
  \bibinfo{author}{\bibfnamefont{T.~G.} \bibnamefont{Cowling}},
  \emph{\bibinfo{title}{The Mathematical Theory of Non-uniform Gases}}
  (\bibinfo{publisher}{Cambridge University Press}, \bibinfo{year}{1970}),
  \bibinfo{edition}{3rd} ed.

\bibitem[{\citenamefont{Jacucci and
  McDonald}(1975)}]{Jacucci1975Structure-and-d}
\bibinfo{author}{\bibfnamefont{G.}~\bibnamefont{Jacucci}} \bibnamefont{and}
  \bibinfo{author}{\bibfnamefont{I.~R.} \bibnamefont{McDonald}},
  \bibinfo{journal}{Physica A} \textbf{\bibinfo{volume}{80}},
  \bibinfo{pages}{607} (\bibinfo{year}{1975}).

\bibitem[{\citenamefont{Tankeshwar and
  Ould-Kaddour}(1992)}]{Tankeshwar1992Tracer-diffusio}
\bibinfo{author}{\bibfnamefont{K.}~\bibnamefont{Tankeshwar}} \bibnamefont{and}
  \bibinfo{author}{\bibfnamefont{F.}~\bibnamefont{Ould-Kaddour}},
  \bibinfo{journal}{J. Phys.: Condens. Matter} \textbf{\bibinfo{volume}{4}},
  \bibinfo{pages}{3349} (\bibinfo{year}{1992}).

\bibitem[{\citenamefont{Sharma and
  Tankeshwar}(1996)}]{Sharma1996Self-diffusion-}
\bibinfo{author}{\bibfnamefont{S.~K.} \bibnamefont{Sharma}} \bibnamefont{and}
  \bibinfo{author}{\bibfnamefont{K.}~\bibnamefont{Tankeshwar}},
  \bibinfo{journal}{J. Phys.: Condens. Matter} \textbf{\bibinfo{volume}{8}},
  \bibinfo{pages}{10839} (\bibinfo{year}{1996}).

\bibitem[{\citenamefont{Tankeshwar et~al.}(1991)\citenamefont{Tankeshwar,
  Singla, and Pathak}}]{Tankeshwar1991A-simple-model-}
\bibinfo{author}{\bibfnamefont{K.}~\bibnamefont{Tankeshwar}},
  \bibinfo{author}{\bibfnamefont{B.}~\bibnamefont{Singla}}, \bibnamefont{and}
  \bibinfo{author}{\bibfnamefont{K.~N.} \bibnamefont{Pathak}},
  \bibinfo{journal}{J. Phys.: Condens. Matter} \textbf{\bibinfo{volume}{3}},
  \bibinfo{pages}{3173} (\bibinfo{year}{1991}).

\bibitem[{\citenamefont{Widom and Rowlinson}(1970)}]{Widom1970New-Model-for-t}
\bibinfo{author}{\bibfnamefont{B.}~\bibnamefont{Widom}} \bibnamefont{and}
  \bibinfo{author}{\bibfnamefont{J.~S.} \bibnamefont{Rowlinson}},
  \bibinfo{journal}{J. Chem. Phys} \textbf{\bibinfo{volume}{52}},
  \bibinfo{pages}{1670} (\bibinfo{year}{1970}).

\bibitem[{\citenamefont{Stillinger and
  Weber}(1978)}]{Stillinger1978Study-of-meltin}
\bibinfo{author}{\bibfnamefont{F.~H.} \bibnamefont{Stillinger}}
  \bibnamefont{and} \bibinfo{author}{\bibfnamefont{T.~A.} \bibnamefont{Weber}},
  \bibinfo{journal}{J. Chem. Phys.} \textbf{\bibinfo{volume}{68}},
  \bibinfo{pages}{3837} (\bibinfo{year}{1978}).

\bibitem[{\citenamefont{Archer and Evans}(2001)}]{Archer2001Binary-Gaussian}
\bibinfo{author}{\bibfnamefont{A.~J.} \bibnamefont{Archer}} \bibnamefont{and}
  \bibinfo{author}{\bibfnamefont{R.}~\bibnamefont{Evans}},
  \bibinfo{journal}{Phys. Rev. E} \textbf{\bibinfo{volume}{64}},
  \bibinfo{pages}{041501} (\bibinfo{year}{2001}).

\bibitem[{\citenamefont{Rapaport}(2004)}]{Rapaport2004The-Art-of-Mole}
\bibinfo{author}{\bibfnamefont{D.~C.} \bibnamefont{Rapaport}},
  \emph{\bibinfo{title}{The Art of Molecular Dynamic Simulation}}
  (\bibinfo{publisher}{Cambridge University Press, Cambridge},
  \bibinfo{year}{2004}), \bibinfo{edition}{2nd} ed.

\bibitem[{\citenamefont{Allen and Tildesley}(1987)}]{Allen1987Computer-Simula}
\bibinfo{author}{\bibfnamefont{M.~P.} \bibnamefont{Allen}} \bibnamefont{and}
  \bibinfo{author}{\bibfnamefont{D.~J.} \bibnamefont{Tildesley}},
  \emph{\bibinfo{title}{Computer Simulations of Liquids}}
  (\bibinfo{publisher}{Oxford University Press, New York},
  \bibinfo{year}{1987}).

\bibitem[{\citenamefont{Shen and Errington}(2006)}]{Shen2006Determination-o}
\bibinfo{author}{\bibfnamefont{V.~K.} \bibnamefont{Shen}} \bibnamefont{and}
  \bibinfo{author}{\bibfnamefont{J.~R.} \bibnamefont{Errington}},
  \bibinfo{journal}{J. Chem. Phys.} \textbf{\bibinfo{volume}{124}},
  \bibinfo{pages}{024721} (\bibinfo{year}{2006}).

\bibitem[{\citenamefont{Shen and Errington}(2005)}]{Shen2005Determination-o}
\bibinfo{author}{\bibfnamefont{V.~K.} \bibnamefont{Shen}} \bibnamefont{and}
  \bibinfo{author}{\bibfnamefont{J.~R.} \bibnamefont{Errington}},
  \bibinfo{journal}{J. Chem. Phys.} \textbf{\bibinfo{volume}{122}},
  \bibinfo{pages}{064508} (\bibinfo{year}{2005}).

\bibitem[{\citenamefont{Boublik}(1970)}]{Boublik1970}
\bibinfo{author}{\bibfnamefont{T.}~\bibnamefont{Boublik}}, \bibinfo{journal}{J.
  Chem. Phys.} \textbf{\bibinfo{volume}{53}}, \bibinfo{pages}{471}
  (\bibinfo{year}{1970}).

\bibitem[{\citenamefont{Monsoori et~al.}(1971)\citenamefont{Monsoori, Carnahan,
  Starling, and Leland}}]{Monsoori1971}
\bibinfo{author}{\bibfnamefont{G.~A.} \bibnamefont{Monsoori}},
  \bibinfo{author}{\bibfnamefont{N.~F.} \bibnamefont{Carnahan}},
  \bibinfo{author}{\bibfnamefont{K.~E.} \bibnamefont{Starling}},
  \bibnamefont{and} \bibinfo{author}{\bibfnamefont{T.~W.}
  \bibnamefont{Leland}}, \bibinfo{journal}{J. Chem. Phys.}
  \textbf{\bibinfo{volume}{54}}, \bibinfo{pages}{1523} (\bibinfo{year}{1971}).

\bibitem[{\citenamefont{Biazzo et~al.}(2009)\citenamefont{Biazzo, Caltagirone,
  Parisi, and Zamponi}}]{Biazzo2009Theory-of-Amorp}
\bibinfo{author}{\bibfnamefont{I.}~\bibnamefont{Biazzo}},
  \bibinfo{author}{\bibfnamefont{F.}~\bibnamefont{Caltagirone}},
  \bibinfo{author}{\bibfnamefont{G.}~\bibnamefont{Parisi}}, \bibnamefont{and}
  \bibinfo{author}{\bibfnamefont{F.}~\bibnamefont{Zamponi}},
  \bibinfo{journal}{Phys. Rev. Lett.} \textbf{\bibinfo{volume}{102}},
  \bibinfo{pages}{195701} (\bibinfo{year}{2009}).

\bibitem[{\citenamefont{Yerazunis et~al.}(1965)\citenamefont{Yerazunis,
  Cornell, and Wintner}}]{Yerazunis1965Dense-Random-Pa}
\bibinfo{author}{\bibfnamefont{S.}~\bibnamefont{Yerazunis}},
  \bibinfo{author}{\bibfnamefont{S.~W.} \bibnamefont{Cornell}},
  \bibnamefont{and} \bibinfo{author}{\bibfnamefont{B.}~\bibnamefont{Wintner}},
  \bibinfo{journal}{Nature} \textbf{\bibinfo{volume}{207}},
  \bibinfo{pages}{835} (\bibinfo{year}{1965}).

\bibitem[{\citenamefont{Hernando}(1990)}]{Hernando1990Thermodynamic-p}
\bibinfo{author}{\bibfnamefont{J.~A.} \bibnamefont{Hernando}},
  \bibinfo{journal}{Mol. Phys.} \textbf{\bibinfo{volume}{69}},
  \bibinfo{pages}{319} (\bibinfo{year}{1990}).

\bibitem[{\citenamefont{Mausbach and May}(2006)}]{Mausbach2006Static-and-dyna}
\bibinfo{author}{\bibfnamefont{P.}~\bibnamefont{Mausbach}} \bibnamefont{and}
  \bibinfo{author}{\bibfnamefont{H.~O.} \bibnamefont{May}},
  \bibinfo{journal}{Fluid Phase Equilib.} \textbf{\bibinfo{volume}{249}},
  \bibinfo{pages}{17} (\bibinfo{year}{2006}).

\bibitem[{\citenamefont{Wensink et~al.}(2008)\citenamefont{Wensink, L\"{o}wen,
  Rex, Likos, and van Teeffelen}}]{Wensink2008Long-time-self-}
\bibinfo{author}{\bibfnamefont{H.}~\bibnamefont{Wensink}},
  \bibinfo{author}{\bibfnamefont{H.}~\bibnamefont{L\"{o}wen}},
  \bibinfo{author}{\bibfnamefont{M.}~\bibnamefont{Rex}},
  \bibinfo{author}{\bibfnamefont{C.}~\bibnamefont{Likos}}, \bibnamefont{and}
  \bibinfo{author}{\bibfnamefont{S.}~\bibnamefont{van Teeffelen}},
  \bibinfo{journal}{Comput. Phys. Commun.} \textbf{\bibinfo{volume}{179}},
  \bibinfo{pages}{77 } (\bibinfo{year}{2008}).

\bibitem[{\citenamefont{Hoyt et~al.}(2000)\citenamefont{Hoyt, Asta, and
  Sadigh}}]{Hoyt2000Test-of-the-Uni}
\bibinfo{author}{\bibfnamefont{J.~J.} \bibnamefont{Hoyt}},
  \bibinfo{author}{\bibfnamefont{M.}~\bibnamefont{Asta}}, \bibnamefont{and}
  \bibinfo{author}{\bibfnamefont{B.}~\bibnamefont{Sadigh}},
  \bibinfo{journal}{Phys. Rev. Lett.} \textbf{\bibinfo{volume}{85}},
  \bibinfo{pages}{594} (\bibinfo{year}{2000}).

\bibitem[{\citenamefont{Cichocki and
  Felderhof}(1990)}]{Cichocki1990Diffusion-coeff}
\bibinfo{author}{\bibfnamefont{B.}~\bibnamefont{Cichocki}} \bibnamefont{and}
  \bibinfo{author}{\bibfnamefont{B.~U.} \bibnamefont{Felderhof}},
  \bibinfo{journal}{J. Chem. Phys.} \textbf{\bibinfo{volume}{93}},
  \bibinfo{pages}{4427} (\bibinfo{year}{1990}).

\bibitem[{\citenamefont{Cichocki and
  Felderhof}(1988)}]{Cichocki1988Long-time-self-}
\bibinfo{author}{\bibfnamefont{B.}~\bibnamefont{Cichocki}} \bibnamefont{and}
  \bibinfo{author}{\bibfnamefont{B.~U.} \bibnamefont{Felderhof}},
  \bibinfo{journal}{J. Chem. Phys} \textbf{\bibinfo{volume}{89}},
  \bibinfo{pages}{3705} (\bibinfo{year}{1988}).

\bibitem[{\citenamefont{Batchelor}(1983)}]{Batchelor1983Diffusion-in-a-}
\bibinfo{author}{\bibfnamefont{G.~K.} \bibnamefont{Batchelor}},
  \bibinfo{journal}{J. Fluid Mech.} \textbf{\bibinfo{volume}{131}},
  \bibinfo{pages}{155} (\bibinfo{year}{1983}).

\bibitem[{\citenamefont{Batchelor}(1976)}]{Batchelor1976Brownian-Diffus}
\bibinfo{author}{\bibfnamefont{G.~K.} \bibnamefont{Batchelor}},
  \bibinfo{journal}{J. Fluid Mech.} \textbf{\bibinfo{volume}{74}},
  \bibinfo{pages}{1} (\bibinfo{year}{1976}).

\bibitem[{\citenamefont{Jeffrey and Onishi}(1984)}]{Jeffrey1984Calculation-of-}
\bibinfo{author}{\bibfnamefont{D.~J.} \bibnamefont{Jeffrey}} \bibnamefont{and}
  \bibinfo{author}{\bibfnamefont{Y.}~\bibnamefont{Onishi}},
  \bibinfo{journal}{J. Fluid Mech.} \textbf{\bibinfo{volume}{139}},
  \bibinfo{pages}{261} (\bibinfo{year}{1984}).

\end{thebibliography}

\end{document}